\documentclass[
aps,pra,
reprint,
a4paper,
floatfix,
longbibliography,
superscriptaddress,
]{revtex4-1}
\usepackage[utf8]{inputenc}
\usepackage[T1]{fontenc}
\usepackage{amsmath,amssymb,amsthm,amsfonts}
\usepackage{mathtools}
\usepackage{braket}
\usepackage{mathrsfs}
\usepackage{hyperref}
\hypersetup{citecolor=magenta}
\hypersetup{colorlinks=true}
\hypersetup{linkcolor=blue}
\hypersetup{urlcolor=blue}

\theoremstyle{definition}
\newtheorem{lemma}{Lemma}

\DeclareMathOperator\tr{tr}
\DeclarePairedDelimiter\abs\lvert\rvert
\DeclarePairedDelimiter\norm\lVert\rVert
\newcommand*\proj[1]{{\ket{#1}\!\bra{#1}}}

\begin{document}

\title{Certifying the activation of Bell nonlocality with finite data}

\author{Jonathan Steinberg}
\affiliation{Naturwissenschaftlich-Technische Fakultät, Universität Siegen, Walter-Flex-Straße 3,
57068 Siegen, Germany}
\affiliation{State Key Laboratory for Mesoscopic Physics, School of Physics and Frontiers Science Center for Nano-Optoelectronics, Peking University, Beijing 100871, China}

\author{H.\ Chau Nguyen}
\email{chau.nguyen@uni-siegen.de}
\affiliation{Naturwissenschaftlich-Technische Fakultät,
Universität Siegen, Walter-Flex-Straße 3, 57068 Siegen, Germany}

\author{Matthias Kleinmann}
\email{matthias.kleinmann@uni-siegen.de}
\affiliation{Naturwissenschaftlich-Technische Fakultät,
Universität Siegen, Walter-Flex-Straße 3, 57068 Siegen, Germany}

\date{\today}
\begin{abstract}
The activation of Bell nonlocality is a protocol that enables the violation of a Bell inequality from a system that initially did not allow for any such violation because the state of the system was Bell-local. This activation of hidden Bell nonlocality has been demonstrated in experiments; however, while the certification of Bell nonlocality is conceptional straightforward, a statistically rigorous verification that the initial state is Bell-local has not yet been achieved. This is due to two key obstacles: The lack of a method to establish a suitable confidence region from the measured data and the need for an efficient technique to verify Bell locality of all states within the confidence region. In this work, we address both challenges. We introduce a confidence polytope in the form of a hyperoctahedron and provide a computationally efficient method to verify whether a quantum state admits a local hidden state model, thus being unsteerable and, consequently, Bell-local. Using these methods, we find that a statistically rigorous certification of hidden Bell nonlocality needs of the order of $10^9$ samples.
\end{abstract}
\maketitle

\section{Introduction}
The measurement of quantum systems distributed among two or more parties can exhibit correlations that carry a signature of the quantum nature of the joint system. Such quantum correlations stratify into various classes. Of particular interest are quantum entanglement, quantum steering, and Bell nonlocality, each referring to a different level of trust in the local measurement devices: In the case of Bell nonlocality one only assumes that the measurements are local, for quantum steering one relies on a trusted quantum description of the measurements of one of the two parties, and for the certification of entanglement one assumes well-characterised quantum measurements for all involved parties. The distinctness of steering and Bell-nonlocal correlations also has been demonstrated in experiments by observing steering correlations from measurements on a state that cannot be used to create Bell-nonlocal correlations \cite{Saunders2010a}.

Interestingly, in certain scenarios, the stratification of quantum correlations can be bypassed through a local filtering process. Then, nonlocal or steering correlations can be activated from quantum states which for themselves would not admit to produce such correlations. This interconversion has been demonstrated in experiments \cite{experiment_kwiat_2001, steerability_superactivation_localmeasure_2019, exp_verification_2020}, but a rigorous statistical analysis of the activation faces difficulties: While it is relatively straightforward to certify the presence of entanglement, quantum steering, or Bell nonlocality by means of measuring an appropriate correlation inequality or entanglement witness, one encounters difficulties in showing that the initial state of the experiment is not already of this class.

Firstly, due to sampling noise and imperfections in the experimental setup, the state can only be determined within a certain error region. Current methods for state tomography \cite{confidence_polytopes_2019, fast_tomography_tropp_2020, comp_conf_regions_almeida_2023} give confidence regions which are either an ellipsoid or a polytope with millions of vertices. Certifying that all states in the confidence region belong to a certain class, such as Bell locality, requires the inspection of every extremal point of the confidence region, and thus cannot be carried out in practice. Addressing this problem, we introduce a simple confidence polytope for quantum state tomography, where the number of vertices scales only linearly with the dimension of the state space and hence quadratically with the dimension of the underlying Hilbert space.

\begin{figure}
    \centering
    \includegraphics[width=75mm,scale=1]{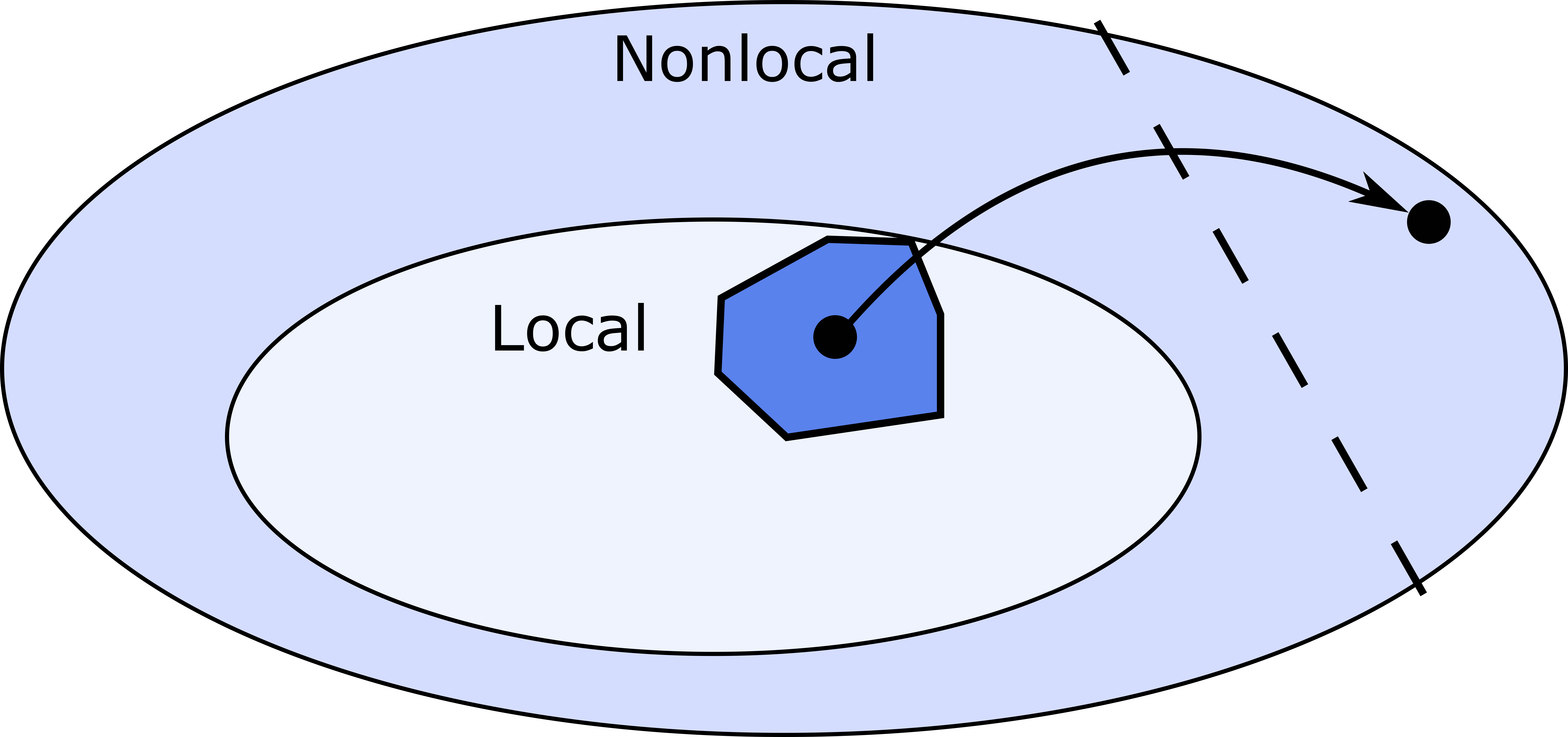}
    \caption{Scheme for demonstrating the activation of Bell nonlocality. Given tomographic measurements of the state, one obtains a corresponding confidence polytope (dark blue area). For each vertex of the polytope one verifies that it cannot violate any Bell inequality and hence is Bell-local. The activation procedure (arrow) uses a filter operation to obtain a Bell-nonlocal state which then can be used to violate a Bell inequality, indicated by the dashed line.}
    \label{fig:eggs}
\end{figure}

Secondly, even for a confidence polytope with a small number of vertices, one still faces the problem that for each vertex one has to decide whether it belongs to the targeted class of correlations. The complexity of this task varies with the class considered: For separable states in low dimensions one can often use criteria like partial transposition \cite{peres_separability_1996, horodecki_separability_1997}, which are computationally efficient. For higher-dimensional entanglement, steering, and Bell-nonlocal states, this problem is computationally much harder and often cannot be solved in reasonable time even for a relatively low number of states. Recently, this problem has received progress for separability \cite{ohst2022a} and Bell locality \cite{hirsch2016, algo_lhv_models_calvacanti_2016, nguyen2018, fillettaz2018, nguyen2019, activation_bound_entanglement_2020}. However, for our use case, the algorithms for Bell-local states are still too slow and yield insufficient accuracy, or are limited to a narrow class of states \cite{villegasaguilar2023nonlocality}. To overcome this difficulty we extend the technique of polytope approximation \cite{nguyen2019} to give a fast and accurate numerical method to solve the case of qudit-qubit systems, which play a crucial role in the activation of Bell nonlocality discussed below.

For the activation of Bell nonlocality, one initially prepares the quantum systems of two distant parties Alice and Bob in a Bell-local state, that is, the state is such that it cannot be used to violate any Bell inequality. But after applying suitable local filters at the distant parties, the transformed state can violate a Bell inequality, see also Figure~\ref{fig:eggs}. The filter consists of local measurements and a communication between both parties where they agree whether they proceed or discard the currently prepared state. An analogous procedure can be formulated for quantum steering inequalities \cite{Saunders2010a, quintino2016}. Our methods provide a statistically conclusive verification of a state being Bell-local and thus allow one to verify the activation of Bell nonlocality and quantum steering. We apply this method to optimise the scenario for experimental feasibility, that is, the number of required number of state preparations in the state tomography.

\section{Activation of Bell nonlocality}
Correlations are Bell-nonlocal if they do not admit a local hidden variable (LHV) model, that is, if the correlations $p(a,b|x,y)$ cannot be described as
\begin{equation}\label{eq:lhv}
    p(a,b|x,y)= \sum_\lambda p(a|x,\lambda)p(b|y,\lambda)p_\lambda,
\end{equation}
where $a$ ($b$) labels the measurement outcome and $x$ ($y$) the measurement setting of Alice (Bob), and $p_\lambda$ is a probability distribution over an inaccessible degree of freedom, that is, over the hidden variable $\lambda$. A Bell inequality bounds the set of correlations achievable by any LHV model and hence, a Bell-local state does not allow Alice and Bob to produce any correlations that cannot be explained by means of an LHV model.

The construction of a LHV model for all measurement settings can, in general, be a difficult problem. Many known LHV models fall into a special more accessible subclass known as local hidden state (LHS) models. The latter describe a weaker form of quantum correlations than Bell nonlocality, known as quantum steering \cite{wiseman2007}. To proceed with a formal definition of a LHS model \cite{wiseman2007, uola2020}, we consider a generalised measurement $x$ on Alice's side, of which the outcomes $a$ are described by effects $E_{a|x}$, that is, by positive semidefinite operators $E_{a|x}\ge 0$ summing up to identity $\sum_{a} E_{a|x} = \openone$. Once the information on the outcome $a$ is available to Bob, his system is ``steered'' to the conditional unnormalized state $\varrho^B_{a|x} = \tr_{A}[\varrho_{\mu,q} E_{a|x} \otimes \openone]$. A LHS model consists now of an ensemble of hidden states on Bob's side, $(p_\lambda,\tau_\lambda)_{\lambda}$, such that all conditional states $\varrho^B_{a|x}$ on his side can be locally simulated. That is, for any generalised measurement $E_{a|x}$, there exists a probability distribution $p(a|x,\lambda)$ such that
\begin{equation}\label{eq:lhs}
    \varrho_{a|x} = \sum_{\lambda} p(a|x,\lambda) p_\lambda \tau_\lambda.
\end{equation}
This means that Bob can explain his conditional states through the preexisting LHSs $(p_\lambda,\tau_\lambda)_{\lambda}$ by assuming that Alice samples from the probability distribution $p(a|x,\lambda)$. Notice that the definition of a LHS model is asymmetric between the parties, leading to the directional nature of quantum steering, in contrast to Bell nonlocality \cite{wiseman2007, uola2020,tischler2018}. Nonetheless, it is straightforward to see that states admitting a LHS model also admit a LHV model~\eqref{eq:lhs} for all measurements \cite{wiseman2007}. In this sense, Bell nonlocality is a stronger form of quantum correlations than quantum steering \cite{wiseman2007}.

\begin{figure}
    \centering
    \includegraphics[width=85mm,scale=0.96]{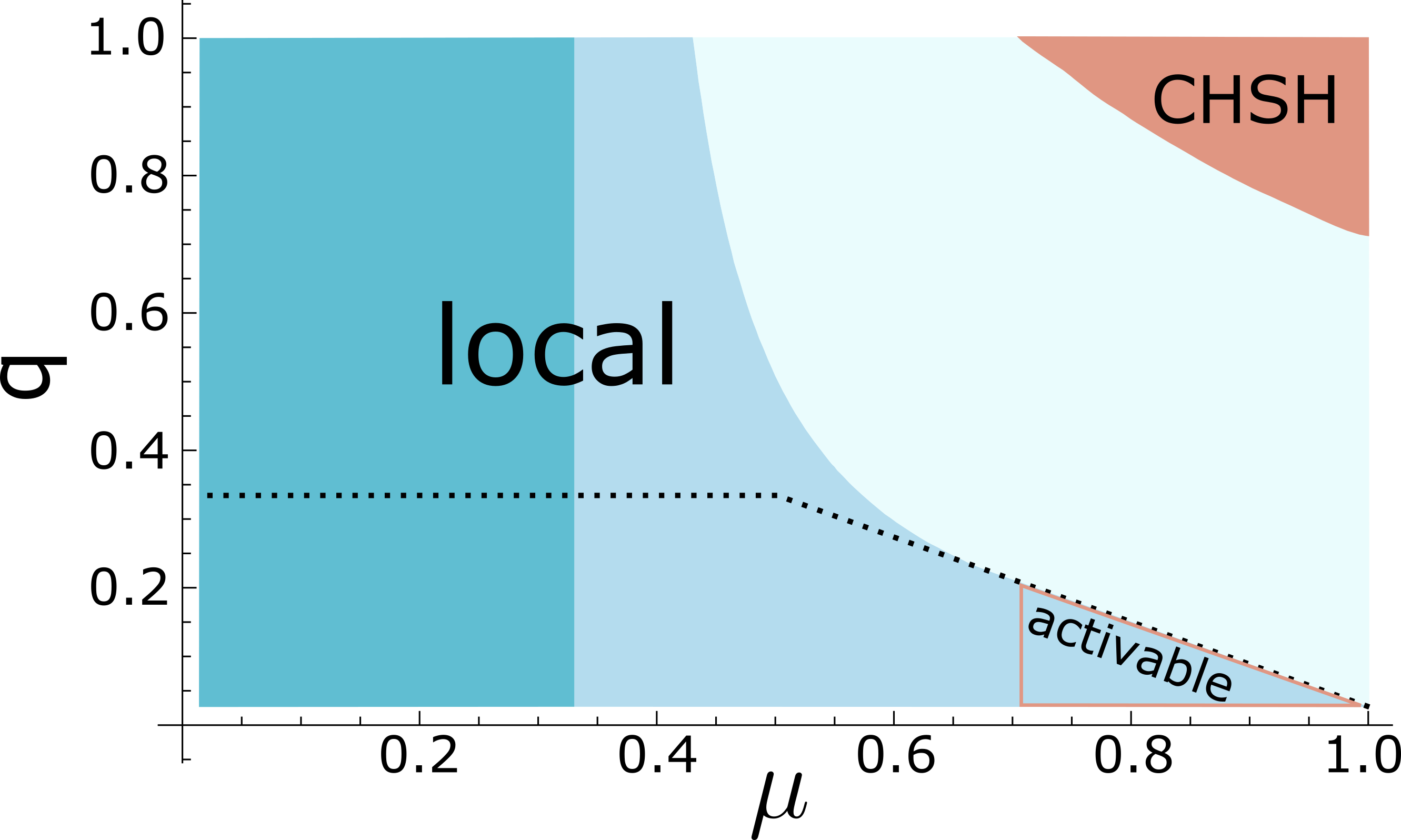}
    \caption{Locality properties of the family of quantum states $\varrho_{\mu,q}$ defined in Eq.~\eqref{eq:nice-state}, depending on the parameters $\mu$ and $q$. The region coloured in dark cyan indicates the range of the parameter $\mu$ for which the Werner state $W_{\mu}$ is separable ($\mu \leq 1/3$) and hence so is $\varrho_{\mu,q}$. The curved, blue region represents the set of parameters for which the state can be proved to be Bell-local. If $\mu,q$ are chosen in the activable region (red triangle), that is $\mu \geq 1/\sqrt{2}$ and $q \leq 2/3(1-\mu)$, $W_{\mu}$ violates the CHSH inequality while the state $\varrho_{\mu,q}$ remains Bell-local. If the parameters are in that region they can be used for the activation of Bell nonlocality. For parameters in the red region (CHSH), the state $\varrho_{\mu,q}$ can violate the CHSH inequality by using appropriate measurements. The Bell locality of the area framed by dotted boundary can be derived from Refs.~\cite{evans2013, baker2018, quintino2015}.
}
    \label{fig:state_family}
\end{figure}

An activation of Bell nonlocality occurs when a state $\varrho$ is Bell-nonlocal but can be brought to violate a Bell inequality after some preprocessing and communication of both parties \cite{experiment_kwiat_2001, steerability_superactivation_localmeasure_2019, exp_verification_2020}. To see the mechanism behind this, let us consider a family of qutrit-qubit states for the system of Alice and Bob,
\begin{align}\label{eq:nice-state}
    \varrho_{\mu,q} = q W_{\mu} + (1-q) \proj2 \otimes \frac{\openone}{2},
\end{align}
with $0 \le \mu \le 1$ and $0 \le q \le 1$. Here, $W_{\mu}$ is the qutrit-qubit embedded Werner state \cite{werner1989},
\begin{equation}
    W_{\mu} = \mu \proj{\psi_-} + (1- \mu) \frac 12(\proj0+\proj1) \otimes \frac\openone2,
\end{equation}
with
\begin{equation}
    \ket{\psi_-} = \frac 1{\sqrt2}(\ket{0}\ket{1}-\ket{1}\ket{0}).
\end{equation}
The state $\varrho_{\mu,q}$ cannot violate any Bell inequality for a certain regime of the parameters $\mu$ and $q$, even when considering generalised measurements \cite{Vertesi2010_TwoQubitBell}. For this we verify that $\varrho_{\mu,q}$ admits a local hidden state (LHS) model, and hence is Bell-local, as described below. Using the results from Refs.~\cite{evans2013, baker2018, quintino2015}, one can show that the state $\varrho_{\mu,q}$ explicitly admits an LHS model for $q$ and $\mu$ in the area framed by the dotted boundary in Figure~\ref{fig:state_family}. We combine this with the results of Ref.~\cite{barrett2002} and construct an LHS model for $q$ and $\mu$ in the whole area labelled by ``local'' in Figure~\ref{fig:state_family}. The detailed derivation and exact description of the corresponding boundary is given in Appendix~\ref{app:rhoprop}. This particularly includes the activable area framed by the solid red triangle $q \le \frac23 (1 - \mu)$ and $1/\sqrt{2} \le \mu \le 1$.

Although the state is Bell-local, Alice and Bob activate its Bell nonlocality by locally filtering the state with (oneway) classical communication. For this, Alice measures the two-outcome measurement $P_\mathrm{keep}=\proj0+\proj1$, $P_\mathrm{discard}=\proj2$ and communicates the result to Bob. If the second outcome occurs, the resulting state is a product state and is discarded by Alice and Bob. Otherwise, Alice and Bob know that they share the Werner state $W_\mu$ and they are ready to perform a Bell test using this state. If $\mu>1/\sqrt{2}$, then the state can violate the Clauser-Horne-Shimony-Holt (CHSH) inequality \cite{chsh_1969}. Therefore, states of the form of Eq.~\eqref{eq:nice-state} which are prepared with parameters $q \le 2/3 ( 1 - \mu)$ and $\mu>1/\sqrt{2} \approx 0.707$ feature hidden Bell nonlocality.

\section{Confidence hyperoctahedron}
\label{sec:tomo}
A conclusive demonstration of the activation of Bell nonlocality must certify that the initial state is Bell-local. This necessitates an accurate description of the joint quantum state of both systems. The most complete picture is obtained from a tomography of the quantum state. For that one uses a collection of measurements where outcome $j$ of setting $k$ is described by the operator $\Pi_{j|k}$. The probabilities $\tr(\Pi_{j|k}\varrho)$ for each outcome are collected into a vector $\vec p$ and the measurements are tomographically complete if $\vec p$ is unique for each state $\varrho$. After measuring each setting for $N$ times, one can then use the resulting vector of observed frequencies $\vec f$ to infer $\hat\varrho$ by inverting $\varrho\mapsto \vec p(\varrho)$.

We require that the dominant source of uncertainty in this process stems from statistical fluctuations, enabling us to perform a statistical hypothesis test and assert whether the physical state is Bell-local with a high degree of confidence. We follow here the strategy to construct a convex confidence region from the data, that is, a region in the state space which contains the experimental state with high probability. It then suffices to verify that all extreme points of this region are Bell-local and hence, by convexity, so are all states in the confidence region. This leads to a low probability for false positives, that is, the test rarely indicates a Bell-local state when the experimental state is Bell-nonlocal.

For this strategy to work, we need the confidence region to have a finite number of extremal points and consequently, the region must be a polytope. Although there are methods available to obtain confidence polytopes \cite{confidence_polytopes_2019}, the huge number of vertices involved makes it computationally infeasible to test whether all of them are Bell-local. Here we use a hyperoctahedron as confidence region where for each dimension of the state space there are only two vertices. The underlying idea is an outer approximation of a confidence ellipsoid in state space, specifically, we use the Gaussian confidence region $C_2$ of Ref.~\cite{comp_conf_regions_almeida_2023} to obtain
\begin{equation}\label{eq:confidence_gauss}
     \mathbb P[\, \norm{ \vec p(\hat \varrho)-\vec p(\varrho) }_2 \le \alpha \,]
     \ge F_\ell(2N\alpha^2).
\end{equation}
This inequality gives a lower bound on the probability that the estimate $\hat\varrho$ deviates from the fiducial state $\varrho$ by at most $\alpha$, where the difference is measured according to the Euclidean norm of the probability vector $\vec p(\varrho)= \tr(\Pi_{j|k}\varrho)_{j,k}$.

More precisely, the estimate $\hat\varrho$ is computed from the observed frequencies $\vec f$ as $\hat\varrho= G(\vec f)+\openone/\tr(\openone)$ with $G$ the pseudo-inverse of the map $X\mapsto \vec p(X)$ on the zero-trace Hermitian operators $X$. The frequencies $\vec f$ are obtained from measuring each setting $k$ for $N$ times and the probability $\mathbb P[\cdots]$ is to be understood with respect to the randomness of the data determining $\vec f$. On the right-hand side, $F_\ell$ denotes the cumulative distribution function of the central $\chi^2(\ell)$ distribution, where $\ell$ is the linear dimension of the state space, that is, $\ell=d^2-1$ for a $d$-dimensional quantum system.

Initially, Eq.~\eqref{eq:confidence_gauss} provides a confidence region in the shape of an ellipsoid. To make this more explicit, we define the ellipsoid $\Theta$ of all zero-trace Hermitian operators $Y$ with $\norm{\vec p(Y)}_2\le 1$. Then $\norm{ \vec p(\hat \varrho)-\vec p(\varrho) }_2\le \alpha$ is equivalent to $\varrho \in \hat\varrho + \alpha\Theta$ and hence, the true state $\varrho$ is with high probability contained in the ellipsoid $\hat\varrho + \alpha\Theta$. We construct an outer approximation of $\Theta$ in the form of a polytope $\Omega$ with vertices $\pm Y_i$ for $i=1,\dotsc \ell$. For this we use that in an $\ell$-dimensional vector space the one-norm unit ball has only $2\ell$ extremal points, and if one scales this ball by $\sqrt \ell$, then it contains the two-norm unit ball. Based on this and replacing $\alpha$ by $\alpha/\sqrt \ell$, we arrive at
\begin{align}\label{eq:poly_confreg}
    \mathbb P[\, \varrho\in \hat \varrho+\alpha\Omega \,]\ge F_\ell\big(\frac{2N}\ell\alpha^2\big),
\end{align}
where $\Omega$ is spanned by $\pm Y_i$ with $(Y_i)_i$ any collection of $\ell$ zero-trace Hermitian operators obeying $\vec p(Y_i)\cdot \vec p(Y_j)=\delta_{i,j}$. A detailed derivation of this confidence hyperoctahedron can be found in Appendix~\ref{app:confpoly}.

\section{Verification of Bell locality}
To demonstrate the Bell locality of the whole confidence polytope, one only needs to determine the Bell locality of all of its vertices. This is, however, a challenging task, as the vertices of the polytope are generally not of a specialised form such as Eq.~\eqref{eq:nice-state} where a special construction of an LHS model is valid \cite{evans2013, baker2018}. A method to construct an LHS model for a generic state $\varrho_{AB}$ is thus required. We address this problem in two steps. First, we reduce the construction of an LHS model for generalised measurements to that of an LHS$_2$ model for dichotomic measurements. Second, we describe how an LHS$_2$ model can be obtained for general qudit-qubit states.

Suppose that for a state $\tilde \varrho_{AB}$ one can construct an LHS$_2$ model. Then it follows \cite{quintino2015, tischler2018} that the state $\frac13 \tilde \varrho_{AB} + \frac23 \sigma_A \otimes \tr_{A}(\tilde \varrho_{AB})$ admits an LHS model for generalised measurements and an arbitrary state $\sigma_A$. Consequently, in order to demonstrate that a state ${\varrho}_{AB}$ admits an LHV model for generalised measurements, it is sufficient that the operator
\begin{equation}
    \tilde{\varrho}_{AB} = 3 {\varrho}_{AB} - 2 \proj 2 \otimes \tr_{A}({\varrho}_{AB})
\end{equation}
admits an LHS$_2$ model. Note that $\tilde \varrho_{AB}$ is not required to be a proper state (that is, not necessarily positive semidefinite) in order for the formal definition of an LHS model to apply, see Appendix~\ref{app:rhoprop}.

Notice that the set of states admitting a LHS$_2$ model is convex \cite{uola2020}. In order to demonstrate that $\tilde{\varrho}_{AB}$ admits an LHS$_2$ model, one defines \cite{nguyen2019}
\begin{multline}
    R(\tilde{\varrho}_{AB}) =\\
    \max\set{t| \tilde{\varrho}_{AB} (t) \text{ admits an LHS$_2$ model} },
\end{multline}
with
\begin{equation}
    \tilde{\varrho}_{AB} (t) =
    t \tilde{\varrho}_{AB} + (1-t) \frac{\openone}{d_A} \otimes \tr_A(\tilde\varrho_{AB}),
\end{equation}
where $d_A$ is Alice's dimension (here $d_A=3$). By virtue of this definition, the state $\tilde{\varrho}_{AB}$ admits an LHS$_2$ model if and only if $R(\tilde{\varrho}_{AB}) \ge 1$ and therefore ${\varrho}_{AB}$ admits an LHS model for generalised measurements.

It has been shown that accurate lower and upper approximations for $R(\tilde{\varrho}_{AB})$ can be computed efficiently for the case of two qubits \cite{nguyen2019}. In this section, we explain how the geometrical method of Ref.~\cite{nguyen2019} can be naturally extended to characterise LHS models over a qudit of dimension $d_A$ at Alice's side coupled to a qubit at Bob's side. We start with approximating Bob's Bloch sphere by a polytope $P$ (from the inside or from the outside), which results in an approximation $R_P(\varrho_{AB})$ for $R(\varrho_{AB})$ (from below and above). The polytope $P$ can be characterised by $N_P$ vertices $\sigma_i$ in the three-dimensional affine space of unit-trace Hermitian operators on Bob's side; we identify $P$ with its vertices $\sigma_i$ and write $P=\set{\sigma_i}_{i=1}^{N_P}$. The summation over the hidden variable $\lambda$ in the LHS model \eqref{eq:lhs} is thus reduced to a finite sum over the vertices of the polytope. The construction of an LHS model is then reduced to finding the probability weights $p_{\lambda}$ with $\lambda$ ranging over the vertices of the approximating polytope such that the response function $p(a|x,\lambda)$ exists for all possible choices of measurements. By formulating the existence of the response function $p(a|x,\lambda)$ as a nesting problem of two convex objects, one can show that the procedure results in a finite linear program for $R_P(\varrho_{AB})$ over $p_{\lambda}$,
\begin{align}
    & \text{maximise:} && t \hspace{.6\linewidth}
    \nonumber\\
    & \text{with respect to:} && t, p_1,p_2,\dotsc,p_{N_P}\ge 0 \nonumber\\
    & \text{such that:} && \label{eq:Rlp}
\end{align}
\begin{align*}
    & \text{(i)} && \sum_\lambda p_\lambda = 1 \\
    & \text{(ii)} && \sum_\lambda p_\lambda \zeta_\lambda(Z) \ge
    t \eta_l (Z) + \frac{l (1 - t)}{d_{A}} \tr (\varrho_{B} Z)\\
\end{align*}
where condition (ii) holds for all $1 \le l \le d_A$ and for all $Z \in \mathcal{F}(P)$. Here, $\zeta_\lambda(Z)=\max\{\left \langle Z,\sigma_\lambda \right \rangle,0\}$ and $\eta_l (Z)$ is the sum over the $l$ maximal eigenvalues of $\tr_B(\varrho_{AB} \openone_{A} \otimes Z)$. Furthermore, $\mathcal{F}(P)$ denotes the set of all operator $Z$ defining a plane that goes through at least three points $(\sigma_{\lambda_1},\sigma_{\lambda_2},\sigma_{\lambda_3})$ of $P$ by $\tr (Z \sigma_{\lambda_i})=0$ for $i=1,2,3$. Notice that $\mathcal{F}(P)$ is completely defined by the polytope $P$, and is independent of the probability weights $p_\lambda$. A full derivation of Eq.~\eqref{eq:Rlp} is given in Appendix~\ref{app:sdp}.

For a polytope of $N_P$ vertices, the program has $\mathcal{O} (N_P)$ variables and $\mathcal{O}(d_A N_P^3)$ constraints, which can be efficiently computed. It is important that the size of the program is only linear in Alice's dimension, rendering the analysis of states such as in Eq.~\eqref{eq:nice-state} feasible. Previous methods based on the approximation of the set of measurements by a polytope \cite{hirsch2016, algo_lhv_models_calvacanti_2016, fillettaz2018, activation_bound_entanglement_2020} lead to a semidefinite program with an exponential size in the approximate polytope. These demand much higher computational resource and cannot be practically applied to our state in Eq.~\eqref{eq:nice-state} with sufficient accuracy.

\section{Experimental feasibility}
The experimental feasibility for the activation of Bell nonlocality depends critically on the number $N'$ of preparations that are needed to certify with high confidence that the initial state $\varrho_{\mu,q}$ is Bell-local. For this we aim to compute the largest scaling factor of the confidence polytope, such that the polytope is still fully contained in the set of Bell-local states. A lower bound on this scaling factor is given by
\begin{align} \label{eq:max_size}
    \epsilon_{\mu,q}^* =
    \min_{\substack{s=\pm1\\i=1\dotsc35}}\max \set{\epsilon| R(\tilde\varrho_{\mu,q} + s \epsilon Y_{i})\ge 1},
\end{align}
where the operators $\pm Y_i$ span a hyperoctahedron with 70 vertices, see Appendix~\ref{app:confpoly}. In Figure~\ref{fig:Figure3}, we display $\epsilon^*_{\mu,q}$ as a function of $\mu$ and $q$. One observes that a large area in the parameter space yields roughly the same maximal value $\epsilon^*\approx 0.001$, rendering the target parameter $q$ and $\mu$ robust to experimental imperfections.

It now follows from the statistical analysis in Section~\ref{sec:tomo} that a sufficient number of 
state preparations is given by $N' = 24\ell F_\ell^{-1}(\gamma) /(2 \epsilon^{*2}_{\mu,q})$, where 
$\gamma$ is the desired level of confidence and $\ell=(d_Ad_B)^2-1=35$ is the linear dimension 
of the state space. The factor of $24$ accounts for the fact that one needs to measure all of the 
$24$ measurement settings $N$ times, and hence the number of state preparations is $N'=24N$. We 
obtain that $N'=7.5\times 10^8$ ($N'=4.6\times 10^8$) state preparations are sufficient for a 
confidence level of ``$3\sigma$'' (``$1\sigma$''), that is, $\gamma= 99.7\%$ ($\gamma=68.3\%$). This 
value can be lowered by incorporating stronger assumptions on Bob's measurement devices: If the 
measurement devices of Bob are assumed to be fully characterised, then an activation of 
Bell-nonlocal correlations is possible for a larger range of the parameters $\mu$ and $q$. 
Correspondingly, instead of a the CHSH inequality, we use a steering inequality, see 
Appendix~\ref{app:activ}. This lowers the required number of preparations to $N'=3.2\times 10^7$ 
($N'=2.0\times 10^7$).

\begin{figure}
    \centering
    \includegraphics[width=85mm,scale=0.85]{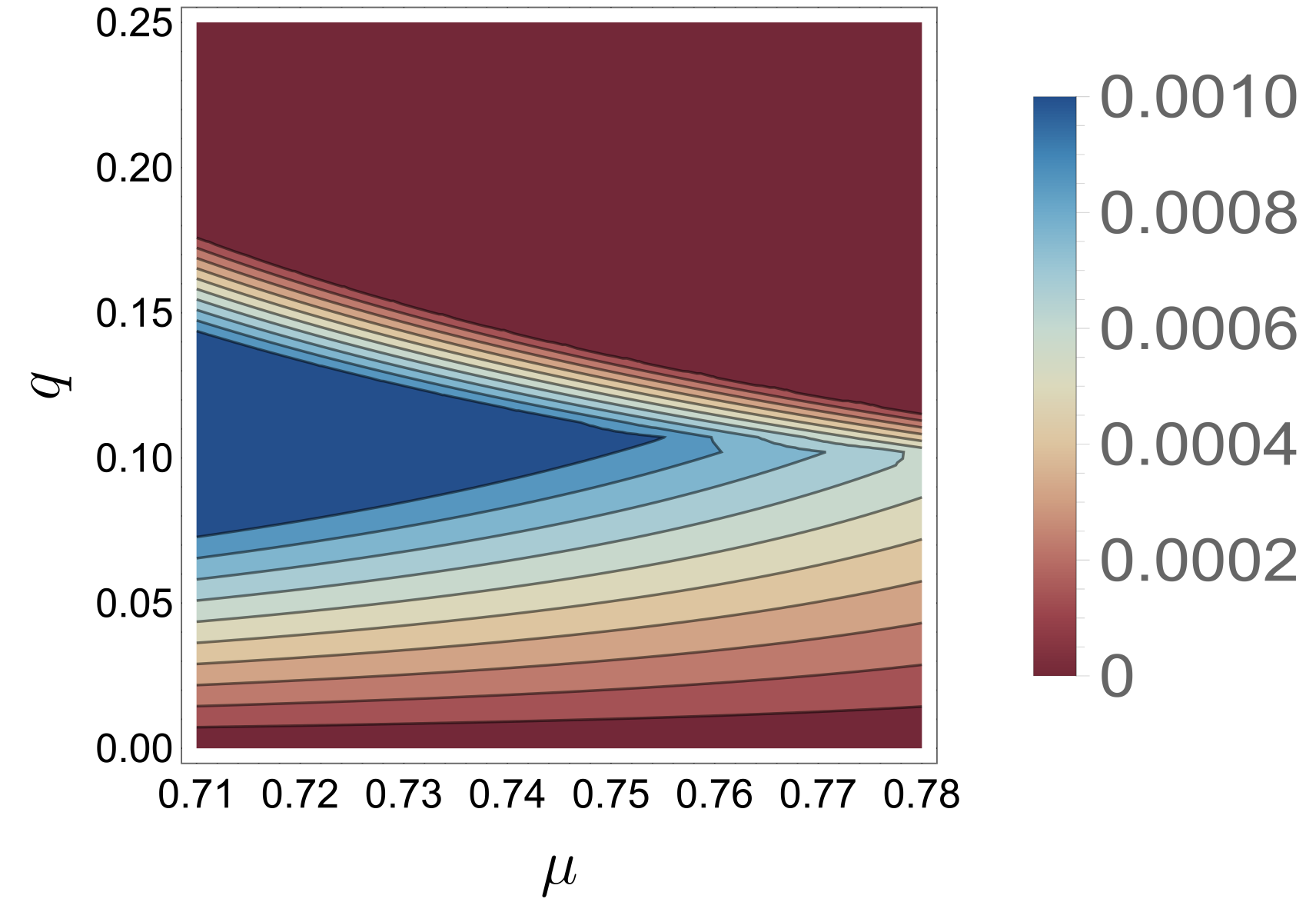}
    \caption{Maximal scaling factor $\epsilon^*_{\mu,q}$ as defined in Eq.~\eqref{eq:max_size} depending on $\mu$ and $q$. All parameter pairs yielding $\epsilon_{\mu,q}^*=0$ (red area) cannot be used for a conclusive activation. As an inner approximation of Bob's Bloch sphere, we have used the icosidodecahedron ($N_P=30$).}
    \label{fig:Figure3}
\end{figure}

\section{Conclusion}
We consider the demonstration of the activation of Bell nonlocality and point out that a rigorous statistical assessment of such an experiment requires: (i) a confidence polytope from state tomography data described by low number of vertices and (ii) a general method to construct local models that can be applied for each of the vertices. To address (i), we derived a confidence region for a quantum state reconstructed from state tomography data in the form of a polytope with the number of vertices scaling quadratically with the system dimension. To address (ii), we generalised the method of polytope approximation designed for two-qubit states \cite{nguyen2018} to the case of a system of qudit and qubit. Specifically for the example of the qutrit-qubit states we consider, we show that $N'=7.5\times 10^8$ state preparations are necessary to obtain a confidence of $\gamma= 99.7\%$ (corresponding to the `$3\sigma$ standard') that the prepared state is local prior to activating.

We notice that both the confidence polytope and the local models have potential applications beyond this specific example. Each of them can be expected to be useful for the statistical analysis of fundamental experiments such as a demonstration of the hierarchy of quantum correlations \cite{Saunders2010a}, the preparation of bound entangled states \cite{amselem_experimental_2009,lavoie_experimental_2010,lavoie_experimental_2010-1} or a demonstration of counter examples to the Peres conjectures \cite{Moroder2014,Vertesi2014}.

\bigskip

\begin{acknowledgments}
The authors would like to thank Ad\'an Cabello, Peter Drmota, Otfried Gühne, Gabriel Araneda Machuca, Fabian Pokorny, Marco Túlio Quintino, Raghavendra Srinivas, Lucas Tendick, Nikolai Wyderka, and Benjamin Yadin for discussions and comments. The University of Siegen is gratefully acknowledged for enabling our computations through the OMNI cluster. This work was supported by the Deutsche Forschungsgemeinschaft (DFG, German Research Foundation, project numbers 447948357 and 440958198), the Sino-German Center for Research Promotion (Project M-0294), the ERC (Consolidator Grant 683107/TempoQ), and the German Ministry of Education and Research (Project QuKuK, BMBF Grant No. 16KIS1618K). JS acknowledges the support from the House of Young Talents of the University of Siegen.
\end{acknowledgments}

\onecolumngrid
\appendix

\section{Locality properties of the family of ideal states}
\label{app:rhoprop}

We want to determine for which values of $0 \le \mu \le 1$, $0 \le q \le 1$ the state
\begin{align}\label{eq:target}
    \varrho_{\mu,q} := q W_{\mu} + (1-q) \vert 2 \rangle \langle 2 \vert \otimes \frac{\openone}{2}
\end{align}
is Bell-local with respect to generalised measurements. Here $W_{\mu}$ is the Werner state \cite{werner1989} only supported on a qubit subspace at Alice's side. Notice that the existence of a local hidden state (LHS) model implies Bell locality. We are going to reduce the construction of an LHS model for generalised measurements to that for dichotomic measurements, making use of the following lemma, first introduced in Ref.~\cite{quintino2015}.
\begin{lemma}[\cite{quintino2015, tischler2018}] \label{lem:extend_quintino}
Suppose $\varrho_{AB} \in B(\mathbb{C}^{d_{A}} \otimes \mathbb{C}^{d_{B}})$ has an LHS model for dichotomic measurements. Then the state
\begin{align}\label{eq:marco_tulio_crit}
    \frac{1}{d_{A}} \varrho_{AB} + \frac{d_{A}-1}{d_{A}} \sigma_{A} \otimes \tr_{A}[\varrho_{AB}]
\end{align}
with arbitrary state $\sigma_{A}$ has an LHS model for arbitrary measurements.
\end{lemma}

Note that the proof of the Lemma in Ref.~\cite{tischler2018} remains valid even when $\varrho_{AB}$ is not positive semidefinite, as long as the conditional states on Bob's side remain positive.

It has been shown that ${\varrho}_{\mu,q}$ admits an LHS model for all dichotomic measurements if $q \le 2 (1-\mu)$ with $1/2 \le \mu \le 1$, or $0 \le \mu \le 1/2$. Using Lemma~\ref{lem:extend_quintino}, one can directly show that the state ${\varrho}_{\mu,q}$ admits an LHS model for all generalised measurements if $q \le 2/3 (1-\mu)$ with $1/2 \le \mu \le 1$, or $q \le 1/3$ with $0 \le \mu \le 1/2$. This forms the area framed by the dotted boundary in Figure~\ref{fig:state_family} in the main text.

\begin{figure}
    \centering
    \includegraphics[width=0.5\textwidth]{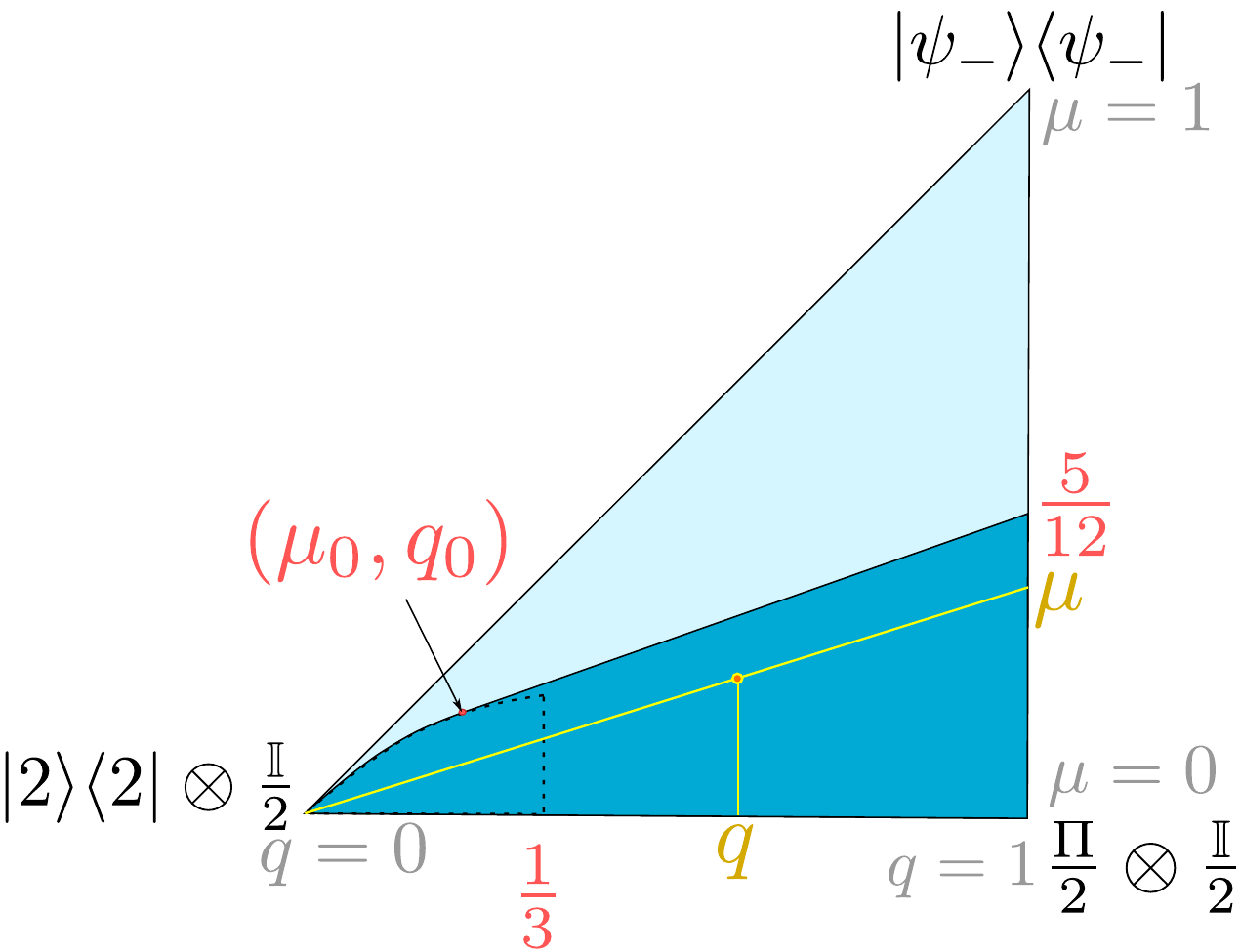}
    \caption{The convex hull of $\Pi \otimes \openone_B$, $\proj{\psi_-}$ and $\proj{2} \otimes 
\openone$ is a triangle in the state space (notice that for the convex geometry, the numerical 
values of the angles are irrelevant). The yellow lines demonstrate how the parameters $q$ and $\mu$ 
for an arbitrary state (orange) in the triangle can be computed. The dotted boundary represents the 
corresponding dotted boundary in Figure~\ref{fig:state_family} in the main text. The convex hull of 
this area with the point ($q=1$, $\mu=5/12$) can be computed by finding the touching point 
$(\mu_0,q_0)$ (red).}
    \label{fig:triangle}
\end{figure}

However, it is known that the Werner state $W_{\mu}$ admits an LHV model with respect to arbitrary generalised measurements for $\mu \leq 5 /12$ \cite{barrett2002}, which is not included in the area framed by dotted boundary in Figure~\ref{fig:state_family} we derived above. The idea then is to consider the convex hull of this area and the point at $(\mu=5/12,q=1)$. Notice that $q$ and $\mu$ parameterise the state space non-linearly. In order to carry out convex geometry operations, we notice that $\varrho_{\mu,q}$ is the convex combination of three points in the state space $\Pi \otimes \openone_B$ (with $\Pi = \proj{0}+\proj{1}$), $\proj{\psi_-}$ (within qubit subspace on Alice's side) and $\proj{2} \otimes \openone$; see Figure~\ref{fig:triangle}. A point in this triangle represents a state $\varrho_{\mu,q}$ and the relation to the parameters $\mu$ and $q$ is illustrated in Figure~\ref{fig:triangle}. The polygonal area in Figure~\ref{fig:state_family} in the main text is now no longer a polygon in Figure~\ref{fig:triangle}. The convex hull of this area is computed by finding a line going through the point corresponding to $\mu=5/12$, $q=1$ that is also a tangent of this area as shown in Figure~\ref{fig:triangle}. To obtain the touching point, it is simplest to transform from $(q,\mu)$ to a linear coordinate system $x=q$, $y = \mu q$. That the touching point satisfies $q = 2/3 (1 -\mu)$ then gives $y = x - 3 x^2/2$. That the tangent at this point goes through $(\mu = 5/12, q =1) \equiv (x_1 = 1, y_1 =5/12)$ gives $1- 3x= (y_1 -y)/(x_1-x)$. A detailed calculation then gives the touching point at $(\mu_0=\sqrt{22}/4-1/2, q_0= 1 - \sqrt{22}/6)$. The linear line connecting $(\mu=\mu_0,q=q_0)$ and $(\mu=5/12,q=1)$ in Figure~\ref{fig:triangle} is translated back to $q \le (-29 + 6 \sqrt{22})/(-24 + 6\sqrt{22} - 12 \mu)$ as plotted in Figure~\ref{fig:state_family} in the main text.

In order to demonstrate an activation of Bell nonlocality, we apply the local filter $F= \vert 2 \rangle \langle 2 \vert$ on Alice's side. Upon measuring $(F,\openone-F)$ and selecting the second outcome, we have the post-measured state
\begin{align}
    \tr[(\openone-F) \varrho_{\mu,q} (\openone-F) ]^{-1}
    (\openone-F) \varrho_{\mu,q}  (\openone-F)  = W_{\mu}.
\end{align}
The Werner state $W_{\mu}$ violates the CHSH inequality \cite{chsh_1969} for $\mu \geq 1 / \sqrt{2}$. We mention that $W_\mu$ is already Bell-nonlocal for smaller values of $\mu$ \cite{LHVmodels_Werner_2017}, but for practical purposes we only consider the case where the filtered state violates the CHSH inequality.

\section{Activation of Steerability}
\label{app:activ}

The applicability of the methods introduced in the main text are not restricted to the activation of Bell nonlocality, where one initially starts with a Bell-local state, that is, a state having an LHS model with respect to \emph{all} generalised measurements, and then certifies that a Bell-nonlocal state was produced by means of local filter operations. Indeed, one could also imagine a scenario where Bob can characterise the states he obtains, yielding more information than just the output statistics of the black-box measurements in the case of CHSH violation. This situation would correspond to the activation of steerability by means of local filters. The crucial point is that the Werner state becomes steerable for smaller parameters $\mu$, that is, it can be steerable but Bell-local.

\begin{figure}
    \centering
    \includegraphics[width=79mm,scale=0.79]{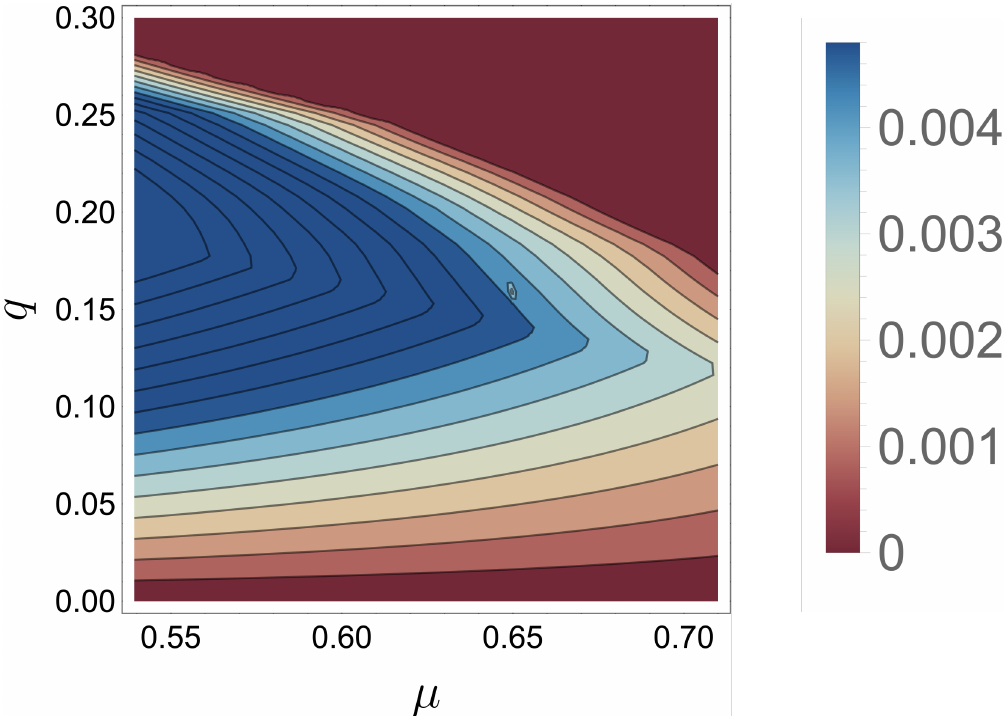}
    \caption{Maximal scaling factor $\epsilon^*_{\mu,q}$ as defined in Eq.~\eqref{eq:max_size} depending on $\mu$ and $q$. Here one allows for smaller parameters $\mu$ as in the main text, as $W_{\mu}$ becomes steerable before it becomes Bell-nonlocal. All parameter pairs yielding $\epsilon_{\mu,q}=0$ (red area) cannot be used for a conclusive activation. As an inner approximation of Bob's Bloch sphere, we have used a icosidodecahedron which has 30 vertices.}
    \label{fig:steering_param}
\end{figure}

The possibility of choosing a smaller $\mu$ yields the existence if larger confidence polytopes in the sense of Eq.~\eqref{eq:max_size} and thus a smaller number of samples to achieve a predefined confidence level. After the filter operation, instead of demonstrating the violation of CHSH inequality, one shows the violation of a so called steering inequality, i.e., one witnesses the non-existence of an LHS model for the filtered state. Motivated by the work in Ref.~\cite{Saunders2010a}, we consider inequalities of the form
\begin{align} \label{eq:steer_inequality}
    S_{N} = \frac{1}{M} \sum_{k=1}^{M} \langle A_{k} \sigma_{\vec{n}_{k}}^{B}\rangle \leq C_{M},
\end{align}
where $A_{k} \in \lbrace \pm 1 \rbrace$ is a random variable describing the outcome of Alice's measurement (still described by a black-box as in the CHSH setting) while Bob explains his outcomes quantum mechanically via the measurement of projective measurements along directions $\vec{n}_{k} \in \mathbb{R}^{3}$. Further, $C_{M}$ denotes the largest value that $S_{M}$ can take when the correlations are explained by means of an LHS model. For the Werner state $W_{\mu}$ it is known that in the limit of $M \rightarrow \infty$ there exist measurement settings for Bob such that $W_{\mu}$ is steerable iff $\mu > 1/2$ \cite{Saunders2010a,wiseman2007}. For the minimal case of $M=2$ measurements, $W_{\mu}$ is steerable iff $\mu \geq 1/\sqrt{2}$ where Bob's measurement directions are given by eigenvectors of $\sigma_{x}$ and $\sigma_{y}$, forming a square on the Bloch sphere. Hence in this setting we do not obtain an decrease of the allowed parameter space of $\mu$. However, if the number of measurements is increased and the directions are chosen in an appropriate manner, the value of $C_{M}$ can be lowered significantly. In fact, it is easy to show that if the state $W_{\mu}$ is prepared, the value of $S_{M}$ in Eq.~\eqref{eq:steer_inequality} is given by $\mu$. If $M=6$ and the measurement directions on Bob's side are given by the octahedron, then one has $C_{6} \approx 0.5393$. Consequently, the parameter regime can be chosen larger, i.e., one can allow $\mu \in [0.5393,1]$ which will offer the possibility for larger confidence polytopes, see Figure~\ref{fig:steering_param}. Indeed, one finds that for $\mu \approx 0.5410$, $q \approx 0.1836$ the maximal size of the polytope is given by $\epsilon \approx 0.0048$, which turns out to be optimal for the allowed parameter region. This implies that the number of samples needed for a conclusive activation of steerability is by a factor $23$ smaller as for the activation of Bell nonlocality explained in the main text. More generally, increasing the number of measurement directions in Eq.~\eqref{eq:steer_inequality} does not lead to a significant smaller number of required samples. For example, with $M=10$ measurement directions the steerability of $W_{\mu}$ can be revealed by $S_{10}$ for $\mu \geq 0.5236$ \cite{Saunders2010a}, a marginal improvement compared to $C_{6}$.

\section{Formulation of steerability with dichotomic measurements as semidefinite program}
\label{app:sdp}

Here we show that deciding whether a quantum state $\varrho_{AB}$ admits an LHS model for dichotomic measurements from Alice to Bob can be solved asymptotically for arbitrary states $\varrho_{AB} \in B(\mathbb{C}^{d} \otimes \mathbb{C}^{2})$. This procedure relies on a reformulation of the problem as a nesting problem of two convex objects \cite{nguyen2018, nguyen2019}. Note that for the most general case of LHS models for generalised measurements, this procedure can in principle also be applied \cite{nguyen2018}, but is practically infeasible as the dimension of the problem grows fast. In general, no efficient method is known to decide the existence of an LHS model for all generalised measurements, even for the case of small dimensions. For that reason, we limit our construction here to dichotomic measurements, and then use Lemma~\ref{lem:extend_quintino} to further construct the LHS model for all generalised measurements.

In order to determine whether an LHS model can be constructed for a quantum state $\varrho_{AB}$ with respect to dichotomic measurements, denoted LHS$_2$ in the main text, one defines the critical radius as
\begin{equation}
    R(\varrho_{AB}) = \max \set{t \ge 0\colon \varrho^t_{AB} \, \text{admits LHS}_2 },
\end{equation}
where $\varrho^t_{AB}=t \varrho_{AB} + (1-t) \frac{\openone_A}{d_A} \otimes \varrho_B $ and $\varrho_B = \tr_{A} (\varrho_{AB})$.
Note that here in the definition we implicitly assumed only dichotomic measurements. It is clear that $\varrho_{AB}$ admits an LHS model if and only if $R(\varrho_{AB}) \ge 1$.

Let $\mu$ be a probability measure over the set of pure states of Bob, denoted by $\mathcal{S}_{B}$. Denote by $\mathcal{K} (\mu)$ the set of all states Alice can simulate with $\mu$, which is referred to as the (simulability) capacity of $\mu$. More precisely,
\begin{equation}
    \mathcal{K} (\mu) 
    = \Set{\int \mathrm{d}\mu (\sigma) g(\sigma) \sigma | g\colon \mathcal S_{B} \to [0,1] }.
\end{equation}
Let $\mathcal M_A$ denote the set of measurement effects of Alice and denote the corresponding conditional state for $E_A\in \mathcal M_A$ by $\tr_A[\varrho_{AB} (E_A \otimes \openone_B)]$. Then $\varrho^{t}_{AB}$ admits an LHS model with respect to dichotomic measurements if the conditional state at Bob's side is a subset of the set of states Alice can simulate for certain $\mu$ \cite{nguyen2018},
\begin{equation}
    \set{\tr_A[\varrho_{AB} (E_A \otimes \openone_B)]|E_A\in\mathcal M_A}
    \subseteq \mathcal{K} (\mu).
\label{eq:nesting}
\end{equation}

\begin{figure}
    \centering
    \includegraphics[width=0.4\textwidth]{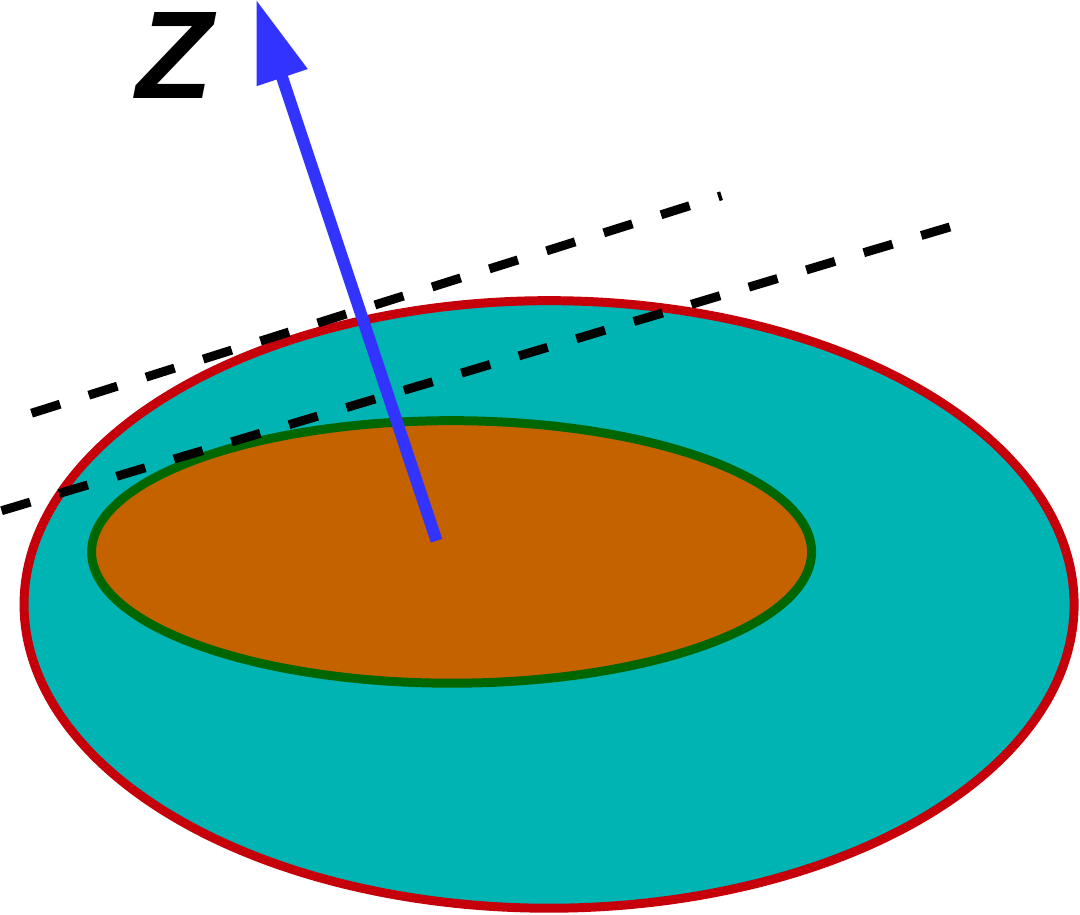}
    \caption{Geometrical intuition for the nesting condition~\eqref{eq:nesting}: Two convex objects are nested if their projections onto any direction are nested.}
    \label{fig:nesting}
\end{figure}

The notion of simulability capacity allows us to rewrite the critical radius as
\begin{align}\begin{split}
    R(\varrho) =\, \,  &\text{max} \, t \\
    &\text{w.r.t.} \, \, t, \mu  \\
    &\text{s.t.} \, \,  \tr [\varrho^{t}_{AB} E_{A} \otimes \openone_{B}] \in \mathcal{K} (\mu)
    \text{ for all } E_A\in \mathcal M_A.
\end{split}\end{align}
Notice that the nesting condition \eqref{eq:nesting} can be expressed as \cite[Lemma~2]{nguyen2018}
\begin{equation}\label{eq:nesting2}
    \max_{\tau \in \mathcal{K}(\mu)} \langle Z_{B},\tau \rangle
    \ge \max_{E_A \in \mathcal{M}_A} \tr (\varrho^{t}_{AB} E_A \otimes Z_B)
\end{equation}
for all Hermitian operators $Z_B$ and with $\braket{Z_B,\tau}=\tr(Z_B\tau)$. This can be interpreted as projecting the nesting~condition~\eqref{eq:nesting2} onto a particular direction define by operator $Z$ in the operator space; see Fig.~\ref{fig:nesting}.

Moreover, using the definition of $\mathcal{K} (\mu)$, one can solve the maximisation on the left hand side, which gives
\begin{equation}
    \max_{\tau \in \mathcal{K} (\mu)} \langle Z_{B}, \tau \rangle = \int \mathrm{d} \mu (\sigma) 
    \max\{\left\langle Z_{B},\sigma \right\rangle,0\}.
\end{equation}

Therefore we obtain the critical radius as
\begin{align}\begin{split}\label{eq:critical2}
    R(\varrho) = \,  &\text{max} \, t \\
    &\text{w.r.t.} \, \,  t, \mu  \\
    &\text{s.t.}  \, \, \int \mathrm{d} \mu (\sigma)
    \max\{\left\langle Z_{B},\sigma \right\rangle,0\}
    \ge \max_{E_A \in \mathcal{M}_A} \tr (\varrho^{t}_{AB} E_A \otimes Z_B)
    \text{ for all } Z_{B}.
\end{split}\end{align}
This is an optimisation over probability measures $\mu$ supported on Bob's Bloch sphere and with respect to an infinite number of constraints, as the whole set of Hermitian operators $Z_{B}$ has to be considered. The crucial step to solve this problem is to introduce a polytope $P$ consisting of $N_P$ vertices, $P=\{\sigma_i\}_{i=1}^{N_P}$, approximating Bob's Bloch sphere $S^B$ from inside or from outside, which gives the lower bound or upper bound of $R(\varrho_{AB})$, respectively. Upon approximating the Bloch sphere by polytope $P$, the probability measure $\mu$ turns into a probability distribution supported on the vertices. We write $u=\{p_k\}_{k=1}^{N_P}$ for this distribution, where $p_{k} = \mu(\sigma_{k})$. Correspondingly, $\mathcal{K}(\mu)$, which can be denoted by $\mathcal{K}_P(u)$ in this case, is also a polytope in the operator space over Bob's system. As a consequence, in the constraints \eqref{eq:critical2}, one only has to consider those $Z^{B}$ corresponding to the normal vectors of the facets of $\mathcal{K}_P (u)$, which are of finite number. Crucially, these normal vectors of $\mathcal{K}_P(u)$ are only dependent on the polytope approximation $P$ of the Bloch sphere, and independent of the probability weights $u$ on the vertices of the polytope.

To find the normal vectors of the facets of the capacity $\mathcal{K}_P(u)$ for a polytope $P$, we follow the following reasoning. As mentioned above, for a certain operator $Z_{B}$, we have
\begin{equation}
    \max_{\tau \in \mathcal{K}_P (u)} \langle Z_{B}, \tau \rangle = \sum_{k=1}^{N_P} p_k  
    \max\{\left\langle Z_{B}, \sigma_k \right\rangle,0\}.
\end{equation}
The operator $Z_{B}$ corresponds to a facet of $\mathcal{K}_P(u)$ if the maximisers on the left form a hyperplane in the four dimensional vector space of operators over Bob's system. The solution on the right hand side gives the maximisers as
\begin{equation}
    \sum_{k \in A} p_k \sigma_k + \sum_{k \in B} p_k \xi_k \sigma_k
\end{equation}
where $A = \{k: \langle Z_{B},\sigma_k \rangle >0\}$ and $B = \{k: \langle Z_{B}, \sigma_k \rangle = 0 \}$ with any $\xi_k$.
One can see that this forms a hyperplane if there are at least $3$ points in $B$, that is, there are three $\sigma_k \in P$ such that $\langle Z_{B}, \sigma_k \rangle = 0$. This means that $P$ defines a plane that goes through at least three points of $P$. Let $\mathcal{F}(P)$ denote the set of all operator $Z_{B}$ that define a plane that goes through at least three points of $P$. Notice that $\mathcal{F}(P)$ is independent of $u$.

However, unlike the problem of two-qubits \cite{nguyen2019}, even under this polytope approximation, the problem \eqref{eq:critical2} is not yet a linear program if Alice owns a qudit. Indeed, the right-hand side of the constraint in \eqref{eq:critical2} still depends on $t$ in a complicated way. However, this complication can be overcome as we will explain in the following. As the objective function is linear, it follows that the maximum is attained at an extreme point of $\mathcal{M}_A$. The extreme points of $\mathcal{M}_A$ are exactly projections of rank-$l$ with $l=0,1,\ldots,d_A$. Consequently,
\begin{equation}
    \max_{E_A \in \mathcal{M}_A} \tr (\varrho^{t}_{AB} E_A \otimes Z^B) = \max_{l=0,1,\ldots,d_A} 
    \max_{E_A \in \mathcal{M}_A, \tr(E_A) = l} \tr (\varrho^{t}_{AB} E_A \otimes Z_B)
\end{equation}
where the right-hand side indeed reduces to a linear expression of $t$,
\begin{equation}
    t \times \max_{E_A \in \mathcal{M}_A, \tr (E_A) = l}
    \tr (\varrho_{AB} E_A \otimes Z_B) + (1 - t) l/d_{A} \tr (\varrho_{B} Z).
\end{equation}
Hence we end up with a linear program of the form
\begin{align}\begin{split}
    R_P(\varrho_{AB}) = \, & \mbox{max} \,  t \\
    &\text{w.r.t.} \,\,  t, \{p_k\}_{k=1}^{N_P}, \sum_{k=1}^{N_P} p_k = 1 \\
    &\text{s.t.} \,  \sum_{k=1}^{N_P} p_k \max\{\left \langle Z_{B},\sigma_k \right \rangle,0\} \ge
    t  \eta_l (Z_{B}) + (1 - t) \frac{l}{d_{A}} \tr (\varrho_{B} Z_{B})\\
    &\quad \text{ for all }l \text{ and for all } Z_{B} \in \mathcal{F}(P)
\end{split}\end{align}
where $\eta_l(Z_{B})= \max \set{\tr (\varrho_{AB} E_A \otimes Z_B)| E_A \in \mathcal{M}_A, \tr(E_A) = l }$. Notice that $\eta_l (Z_{B})$ is simply the sum over the $l$ maximal eigenvalues of $\tr_B(\varrho_{AB} \openone_{A} \otimes Z_B)$.

\section{Construction of the confidence polytope}
\label{app:confpoly}

Quantum state tomography is a method to identify the density operator $\varrho$ of a quantum system by performing measurements across a tomographically complete set of settings. The outcome $j$ of setting $k$ is associated with a positive semidefinite operator $\Pi_{j|k}$ where for each setting one has $\sum_j \Pi_{j|k}= \openone$. The probability of obtaining outcome $j$ when measuring setting $k$ is given by $\tr(\Pi_{j|k}\varrho)$ and the measurement settings are tomographically complete if the collection of all probabilities $\vec p(\varrho)=(\tr(\Pi_{j|k}\varrho))_{j,k}$ is sufficient to identify the state $\varrho$.

In a tomography experiment, each setting is measured $N$ times yielding the relative frequencies $\vec f$ distributed according to $\vec p$. From the frequencies $\vec f$ one computes the estimate \cite{comp_conf_regions_almeida_2023}
\begin{equation}
    \hat\varrho = G(\vec f) + \frac \openone d,
\end{equation}
where $G$ denotes the pseudo-inverse of the linear map $X\mapsto \vec p(X)$ on the zero-trace Hermitian operators $X$ and $d=\tr(\openone)$ is the dimension of the quantum system. Using a Gaussian approximation for the  distribution of $\vec f$, one finds Eq.~\eqref{eq:confidence_gauss} of the main text,
\begin{equation}\label{eq:app:cg}
     \mathbb P[\, \norm{\vec p(\hat \varrho)-\vec p(\varrho)}_2\le\alpha \,]\ge F_\ell(2N\alpha^2),
\end{equation}
see Ref.~\cite{comp_conf_regions_almeida_2023}. By introducing the ellipsoid $\Theta= \set{X | \norm{\vec p(X)}_2\le 1}$ in the space of zero-trace Hermitian operators, we can rewrite the condition $\norm{\vec p(\hat \varrho)-\vec p(\varrho)}_2\le\alpha$ as $\varrho\in \hat \varrho +\alpha\Theta$. Hence, the ellipsoid $\hat \varrho +\alpha\Theta$ constitutes a confidence region once we choose $\alpha$ according to our intended level of confidence (for example, $\gamma=99\%$), that is, we obtain $\alpha$ by solving $F_\ell(2N\alpha^2)=\gamma$.

\begin{figure}
    \centering
    \includegraphics[width=60mm,scale=0.6]{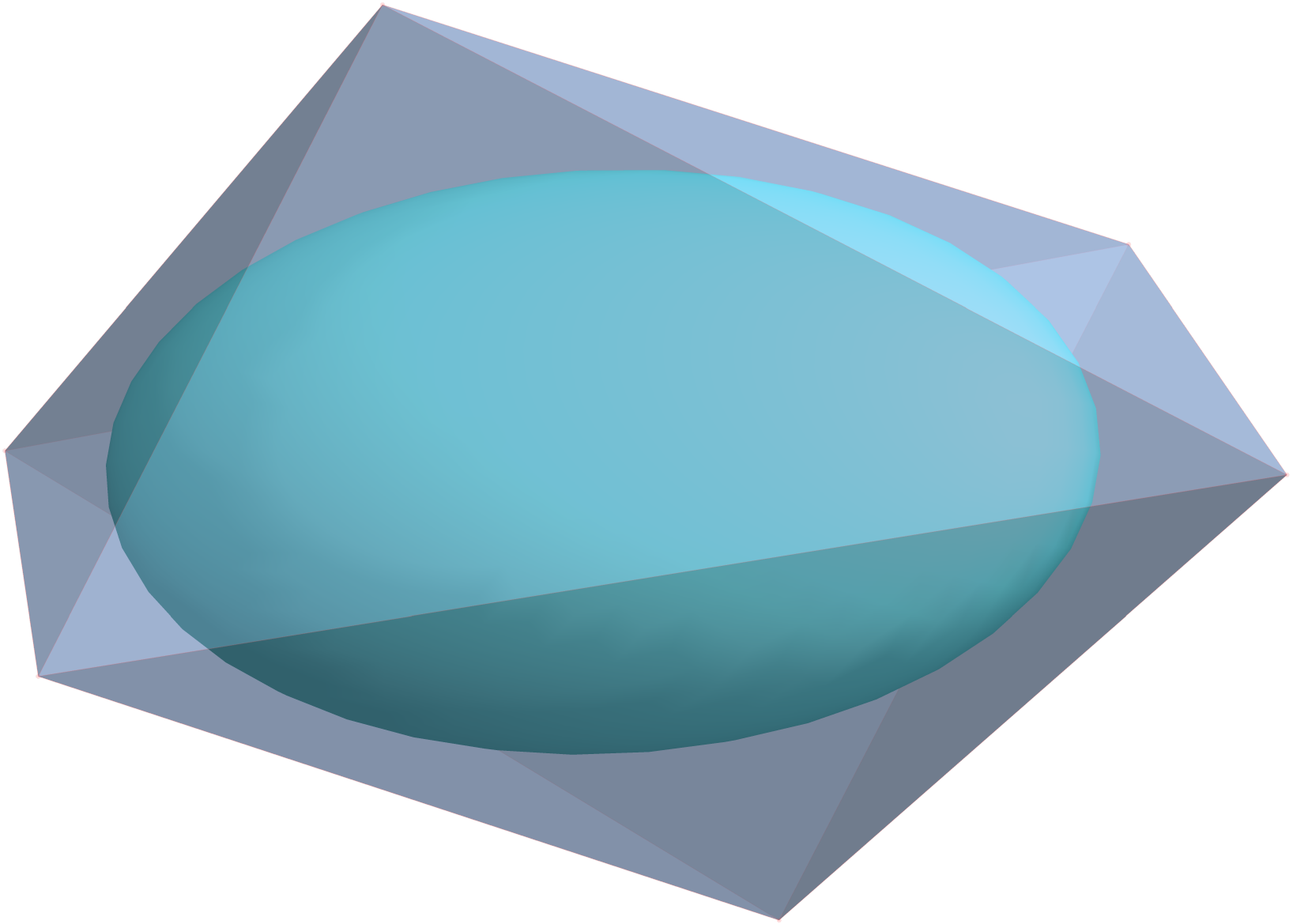}
    \caption{Illustration of the outer approximation of a confidence region by a polytope. The inner ellipsoid is the confidence region in state space. The outer polytope corresponds to the outer approximation by an octahedron.}
    \label{fig:polysphere}
\end{figure}

In order to obtain an outer approximation of the confidence ellipsoid in the form of a hyperoctahedron, we replace the two-norm in the definition of the ellipsoid by the one-norm. For this we use that in an $\ell$-dimensional vector space with orthonormal basis $(\vec b_i)_i$, we have
\begin{equation}
    \norm{\vec v}_2 = \sqrt{\sum_i \abs{\vec b_i \cdot \vec v}^2}
    \ge \frac1{\sqrt \ell} \sum_i \abs{\vec b_i\cdot \vec v}.
\end{equation}
The two-norm unit ball $\set{\vec v| \norm{\vec v}_2\le 1 }$ is hence contained in the hyperoctahedron $\set{\sqrt \ell\,\vec v|\sum_i \abs{\vec b_i\cdot \vec v}\le 1}$ with extremal points $\pm\sqrt \ell \,\vec b_i$, see Figure~\ref{fig:polysphere} for an illustration. We apply this to the $\ell$-dimensional range of $X\mapsto \vec p(X)$ over all zero-trace Hermitian operators $X$. Any orthonormal basis $\vec b_i$ of this range can be written as $\vec b_i=\vec p(Y_i)$ with $Y_i$ appropriate zero-trace Hermitian operators. This gives us
\begin{equation}
    \Theta = \Set{ G[\vec p(X)]| \sum_i [\,\vec p(Y_i)\cdot \vec p(X)\,]^2\le 1}
    \subset \sqrt \ell\, \mathrm{conv}\set{\pm G[\vec p(Y_i)]| i} = \sqrt \ell\, \Omega
\end{equation}
with $\Omega=\mathrm{conv}\set{\pm Y_i| i}$ the convex hull of all $\pm Y_i$. After rescaling $\alpha$, we obtain Eq.~\eqref{eq:poly_confreg} of the main text,
\begin{equation}
    \mathbb P[\, \varrho\in \hat \varrho+\alpha\Omega ]\ge F_\ell\big(\frac{2N}\ell\alpha^2\big).
\end{equation}

\medskip

For our specific scenario of one qutrit and one qubit we choose as measurement observables the Pauli operators $\sigma_x$, $\sigma_y$, $\sigma_z$ for the qubit and the operators $\sigma_{0,x}$, $\sigma_{0,y}$, $\sigma_{1,x}$, $\sigma_{1,y}$, $\sigma_{1,z}$, $\sigma_{2,x}$, $\sigma_{2,y}$, $\sigma_{2,z}$, for the qutrit, where $\sigma_{k,\mu}$ acts like $\sigma_\mu$ on the two-dimensional subspace orthogonal to $\ket k$ and $\sigma_{k,\mu}\ket k=\ket k$, for example,
\begin{equation}
    \braket{i|\sigma_{1,y}|j}=\begin{pmatrix}0&0&-i\\0&1&0\\i&0&0\end{pmatrix}_{i,j}.
\end{equation}
Hence, we have 24 measurement settings, each with 4 outcomes, corresponding to the eigenvalues $\pm1$ for the qutrit and qubit observables, each. We omit the outcome from our analysis, where both measurements yield $-1$, which reduces the vector $\vec p$ to have only $72$ entries. Then one readily chooses a basis $(Y_i)_i$ of the zero-trace Hermitian operators, such that $\vec p(Y_i)\cdot \vec p(Y_j)=\delta_{i,j}$ holds. Since the linear dimension of the zero-trace Hermitian operators is $\ell=d_Ad_B-1=35$ and $\vec p$ is linear, exactly $2\ell$ operators $\pm Y_1,\dotsc,\pm Y_{35}$ span the hyperoctahedron described above. There remains a freedom in the orientation of the hyperoctahedron since the operators $Y_i$ only need to satisfy the aforementioned orthogonality conditions. We choose the operators at random, such the orientation of the polytope does not prefer any specific direction. The specific choice of the operators $Y_i$ for our numerical analysis is provided in a machine-readable format upon request.

\twocolumngrid

\bibliography{ref.bib}

\begin{thebibliography}{34}%
\makeatletter
\providecommand \@ifxundefined [1]{%
 \@ifx{#1\undefined}
}%
\providecommand \@ifnum [1]{%
 \ifnum #1\expandafter \@firstoftwo
 \else \expandafter \@secondoftwo
 \fi
}%
\providecommand \@ifx [1]{%
 \ifx #1\expandafter \@firstoftwo
 \else \expandafter \@secondoftwo
 \fi
}%
\providecommand \natexlab [1]{#1}%
\providecommand \enquote  [1]{``#1''}%
\providecommand \bibnamefont  [1]{#1}%
\providecommand \bibfnamefont [1]{#1}%
\providecommand \citenamefont [1]{#1}%
\providecommand \href@noop [0]{\@secondoftwo}%
\providecommand \href [0]{\begingroup \@sanitize@url \@href}%
\providecommand \@href[1]{\@@startlink{#1}\@@href}%
\providecommand \@@href[1]{\endgroup#1\@@endlink}%
\providecommand \@sanitize@url [0]{\catcode `\\12\catcode `\$12\catcode
  `\&12\catcode `\#12\catcode `\^12\catcode `\_12\catcode `\%12\relax}%
\providecommand \@@startlink[1]{}%
\providecommand \@@endlink[0]{}%
\providecommand \url  [0]{\begingroup\@sanitize@url \@url }%
\providecommand \@url [1]{\endgroup\@href {#1}{\urlprefix }}%
\providecommand \urlprefix  [0]{URL }%
\providecommand \Eprint [0]{\href }%
\providecommand \doibase [0]{http://dx.doi.org/}%
\providecommand \selectlanguage [0]{\@gobble}%
\providecommand \bibinfo  [0]{\@secondoftwo}%
\providecommand \bibfield  [0]{\@secondoftwo}%
\providecommand \translation [1]{[#1]}%
\providecommand \BibitemOpen [0]{}%
\providecommand \bibitemStop [0]{}%
\providecommand \bibitemNoStop [0]{.\EOS\space}%
\providecommand \EOS [0]{\spacefactor3000\relax}%
\providecommand \BibitemShut  [1]{\csname bibitem#1\endcsname}%
\let\auto@bib@innerbib\@empty
%</preamble>
\bibitem [{\citenamefont {Saunders}\ \emph {et~al.}(2010)\citenamefont
  {Saunders}, \citenamefont {Jones}, \citenamefont {Wiseman},\ and\
  \citenamefont {Pryde}}]{Saunders2010a}%
  \BibitemOpen
  \bibfield  {author} {\bibinfo {author} {\bibfnamefont {D.~J.}\ \bibnamefont
  {Saunders}}, \bibinfo {author} {\bibfnamefont {S.~J.}\ \bibnamefont {Jones}},
  \bibinfo {author} {\bibfnamefont {H.~M.}\ \bibnamefont {Wiseman}}, \ and\
  \bibinfo {author} {\bibfnamefont {G.~J.}\ \bibnamefont {Pryde}},\ }\bibfield
  {title} {\enquote {\bibinfo {title} {Experimental {EPR}-steering using
  {Bell}-local states},}\ }\href {\doibase 10.1038/nphys1766} {\bibfield
  {journal} {\bibinfo  {journal} {Nat. Phys.}\ }\textbf {\bibinfo {volume}
  {6}},\ \bibinfo {pages} {845} (\bibinfo {year} {2010})}\BibitemShut {NoStop}%
\bibitem [{\citenamefont {Kwiat}\ \emph {et~al.}(2001)\citenamefont {Kwiat},
  \citenamefont {Barraza-Lopez}, \citenamefont {Stefanov},\ and\ \citenamefont
  {Gisin}}]{experiment_kwiat_2001}%
  \BibitemOpen
  \bibfield  {author} {\bibinfo {author} {\bibfnamefont {P.~G.}\ \bibnamefont
  {Kwiat}}, \bibinfo {author} {\bibfnamefont {S.}~\bibnamefont
  {Barraza-Lopez}}, \bibinfo {author} {\bibfnamefont {A.}~\bibnamefont
  {Stefanov}}, \ and\ \bibinfo {author} {\bibfnamefont {N.}~\bibnamefont
  {Gisin}},\ }\bibfield  {title} {\enquote {\bibinfo {title} {Experimental
  entanglement distillation and ``hidden'' non-locality},}\ }\href {\doibase
  10.1038/35059017} {\bibfield  {journal} {\bibinfo  {journal} {Nature}\
  }\textbf {\bibinfo {volume} {409}},\ \bibinfo {pages} {1014} (\bibinfo {year}
  {2001})}\BibitemShut {NoStop}%
\bibitem [{\citenamefont {Pramanik}\ \emph {et~al.}(2019)\citenamefont
  {Pramanik}, \citenamefont {Cho}, \citenamefont {Han}, \citenamefont {Lee},
  \citenamefont {Kim},\ and\ \citenamefont
  {Moon}}]{steerability_superactivation_localmeasure_2019}%
  \BibitemOpen
  \bibfield  {author} {\bibinfo {author} {\bibfnamefont {T.}~\bibnamefont
  {Pramanik}}, \bibinfo {author} {\bibfnamefont {Y.~W.}\ \bibnamefont {Cho}},
  \bibinfo {author} {\bibfnamefont {S.~W.}\ \bibnamefont {Han}}, \bibinfo
  {author} {\bibfnamefont {S.~Y.}\ \bibnamefont {Lee}}, \bibinfo {author}
  {\bibfnamefont {Y.~Su}\ \bibnamefont {Kim}}, \ and\ \bibinfo {author}
  {\bibfnamefont {S.}~\bibnamefont {Moon}},\ }\bibfield  {title} {\enquote
  {\bibinfo {title} {Revealing hidden quantum steerability using local
  filtering operations},}\ }\href {\doibase 10.1103/PhysRevA.99.030101}
  {\bibfield  {journal} {\bibinfo  {journal} {Phys. Rev. A}\ }\textbf {\bibinfo
  {volume} {99}},\ \bibinfo {pages} {030101} (\bibinfo {year}
  {2019})}\BibitemShut {NoStop}%
\bibitem [{\citenamefont {Wang}\ \emph {et~al.}(2020)\citenamefont {Wang},
  \citenamefont {Li}, \citenamefont {Wang}, \citenamefont {Liu},\ and\
  \citenamefont {Wang}}]{exp_verification_2020}%
  \BibitemOpen
  \bibfield  {author} {\bibinfo {author} {\bibfnamefont {Y.}~\bibnamefont
  {Wang}}, \bibinfo {author} {\bibfnamefont {J.}~\bibnamefont {Li}}, \bibinfo
  {author} {\bibfnamefont {X.~R.}\ \bibnamefont {Wang}}, \bibinfo {author}
  {\bibfnamefont {T.~J.}\ \bibnamefont {Liu}}, \ and\ \bibinfo {author}
  {\bibfnamefont {Q.}~\bibnamefont {Wang}},\ }\bibfield  {title} {\enquote
  {\bibinfo {title} {Experimental demonstration of hidden nonlocality with
  local filters},}\ }\href {\doibase 10.1364/OE.387568} {\bibfield  {journal}
  {\bibinfo  {journal} {Opt. Express}\ }\textbf {\bibinfo {volume} {28}},\
  \bibinfo {pages} {13638} (\bibinfo {year} {2020})}\BibitemShut {NoStop}%
\bibitem [{\citenamefont {Wang}\ \emph {et~al.}(2019)\citenamefont {Wang},
  \citenamefont {Scholz},\ and\ \citenamefont
  {Renner}}]{confidence_polytopes_2019}%
  \BibitemOpen
  \bibfield  {author} {\bibinfo {author} {\bibfnamefont {J.}~\bibnamefont
  {Wang}}, \bibinfo {author} {\bibfnamefont {V.~B.}\ \bibnamefont {Scholz}}, \
  and\ \bibinfo {author} {\bibfnamefont {R.}~\bibnamefont {Renner}},\
  }\bibfield  {title} {\enquote {\bibinfo {title} {Confidence polytopes in
  quantum state tomography},}\ }\href {\doibase 10.1103/PhysRevLett.122.190401}
  {\bibfield  {journal} {\bibinfo  {journal} {Phys. Rev. Lett.}\ }\textbf
  {\bibinfo {volume} {122}},\ \bibinfo {pages} {190401} (\bibinfo {year}
  {2019})}\BibitemShut {NoStop}%
\bibitem [{\citenamefont {Guţă}\ \emph {et~al.}(2020)\citenamefont {Guţă},
  \citenamefont {Kahn}, \citenamefont {Kueng},\ and\ \citenamefont
  {Tropp}}]{fast_tomography_tropp_2020}%
  \BibitemOpen
  \bibfield  {author} {\bibinfo {author} {\bibfnamefont {M.}~\bibnamefont
  {Guţă}}, \bibinfo {author} {\bibfnamefont {J.}~\bibnamefont {Kahn}},
  \bibinfo {author} {\bibfnamefont {R.}~\bibnamefont {Kueng}}, \ and\ \bibinfo
  {author} {\bibfnamefont {J.~A.}\ \bibnamefont {Tropp}},\ }\bibfield  {title}
  {\enquote {\bibinfo {title} {Fast state tomography with optimal error
  bounds},}\ }\href {\doibase 10.1088/1751-8121/ab8111} {\bibfield  {journal}
  {\bibinfo  {journal} {J. Phys. A: Math. Theor.}\ }\textbf {\bibinfo {volume}
  {53}},\ \bibinfo {pages} {204001} (\bibinfo {year} {2020})}\BibitemShut
  {NoStop}%
\bibitem [{\citenamefont {de~Almeida}\ \emph {et~al.}(2023)\citenamefont
  {de~Almeida}, \citenamefont {Kleinmann},\ and\ \citenamefont
  {Sentís}}]{comp_conf_regions_almeida_2023}%
  \BibitemOpen
  \bibfield  {author} {\bibinfo {author} {\bibfnamefont {J.~O.}\ \bibnamefont
  {de~Almeida}}, \bibinfo {author} {\bibfnamefont {M.}~\bibnamefont
  {Kleinmann}}, \ and\ \bibinfo {author} {\bibfnamefont {G.}~\bibnamefont
  {Sentís}},\ }\bibfield  {title} {\enquote {\bibinfo {title} {Comparison of
  confidence regions for quantum state tomography},}\ }\href {\doibase
  10.1088/1367-2630/ad06d9} {\bibfield  {journal} {\bibinfo  {journal} {New J.
  Phys.}\ }\textbf {\bibinfo {volume} {25}},\ \bibinfo {pages} {113018}
  (\bibinfo {year} {2023})}\BibitemShut {NoStop}%
\bibitem [{\citenamefont {Peres}(1996)}]{peres_separability_1996}%
  \BibitemOpen
  \bibfield  {author} {\bibinfo {author} {\bibfnamefont {A.}~\bibnamefont
  {Peres}},\ }\bibfield  {title} {\enquote {\bibinfo {title} {Separability
  criterion for density matrices},}\ }\href {\doibase
  10.1103/PhysRevLett.77.1413} {\bibfield  {journal} {\bibinfo  {journal}
  {Phys. Rev. Lett.}\ }\textbf {\bibinfo {volume} {77}},\ \bibinfo {pages}
  {1413} (\bibinfo {year} {1996})}\BibitemShut {NoStop}%
\bibitem [{\citenamefont {Horodecki}(1997)}]{horodecki_separability_1997}%
  \BibitemOpen
  \bibfield  {author} {\bibinfo {author} {\bibfnamefont {P.}~\bibnamefont
  {Horodecki}},\ }\bibfield  {title} {\enquote {\bibinfo {title} {Separability
  criterion and inseparable mixed states with positive partial
  transposition},}\ }\href {\doibase 10.1016/S0375-9601(97)00416-7} {\bibfield
  {journal} {\bibinfo  {journal} {Phys. Lett. A}\ }\textbf {\bibinfo {volume}
  {232}},\ \bibinfo {pages} {333} (\bibinfo {year} {1997})}\BibitemShut
  {NoStop}%
\bibitem [{\citenamefont {Ohst}\ \emph {et~al.}(2024)\citenamefont {Ohst},
  \citenamefont {Yu}, \citenamefont {Gühne},\ and\ \citenamefont
  {Nguyen}}]{ohst2022a}%
  \BibitemOpen
  \bibfield  {author} {\bibinfo {author} {\bibfnamefont {Ties-A.}\ \bibnamefont
  {Ohst}}, \bibinfo {author} {\bibfnamefont {Xiao-Dong}\ \bibnamefont {Yu}},
  \bibinfo {author} {\bibfnamefont {Otfried}\ \bibnamefont {Gühne}}, \ and\
  \bibinfo {author} {\bibfnamefont {H.~Chau}\ \bibnamefont {Nguyen}},\
  }\bibfield  {title} {\enquote {\bibinfo {title} {Certifying quantum
  separability with adaptive polytopes},}\ }\href {\doibase
  10.21468/SciPostPhys.16.3.063} {\bibfield  {journal} {\bibinfo  {journal}
  {SciPost Phys.}\ }\textbf {\bibinfo {volume} {16}},\ \bibinfo {pages} {063}
  (\bibinfo {year} {2024})}\BibitemShut {NoStop}%
\bibitem [{\citenamefont {Hirsch}\ \emph {et~al.}(2016)\citenamefont {Hirsch},
  \citenamefont {Quintino}, \citenamefont {V\'ertesi}, \citenamefont {Pusey},\
  and\ \citenamefont {Brunner}}]{hirsch2016}%
  \BibitemOpen
  \bibfield  {author} {\bibinfo {author} {\bibfnamefont {F.}~\bibnamefont
  {Hirsch}}, \bibinfo {author} {\bibfnamefont {M.~T.}\ \bibnamefont
  {Quintino}}, \bibinfo {author} {\bibfnamefont {T.}~\bibnamefont {V\'ertesi}},
  \bibinfo {author} {\bibfnamefont {M.~F.}\ \bibnamefont {Pusey}}, \ and\
  \bibinfo {author} {\bibfnamefont {N.}~\bibnamefont {Brunner}},\ }\bibfield
  {title} {\enquote {\bibinfo {title} {Algorithmic construction of local hidden
  variable models for entangled quantum states},}\ }\href {\doibase
  10.1103/PhysRevLett.117.190402} {\bibfield  {journal} {\bibinfo  {journal}
  {Phys. Rev. Lett.}\ }\textbf {\bibinfo {volume} {117}},\ \bibinfo {pages}
  {190402} (\bibinfo {year} {2016})}\BibitemShut {NoStop}%
\bibitem [{\citenamefont {Calvacanti}\ \emph {et~al.}(2016)\citenamefont
  {Calvacanti}, \citenamefont {Guerini}, \citenamefont {Rabelo},\ and\
  \citenamefont {Skrzypczyk}}]{algo_lhv_models_calvacanti_2016}%
  \BibitemOpen
  \bibfield  {author} {\bibinfo {author} {\bibfnamefont {D.}~\bibnamefont
  {Calvacanti}}, \bibinfo {author} {\bibfnamefont {L.}~\bibnamefont {Guerini}},
  \bibinfo {author} {\bibfnamefont {R.}~\bibnamefont {Rabelo}}, \ and\ \bibinfo
  {author} {\bibfnamefont {P.}~\bibnamefont {Skrzypczyk}},\ }\bibfield  {title}
  {\enquote {\bibinfo {title} {General method for constructing local hidden
  variable models for entangled quantum states},}\ }\href {\doibase
  10.1103/PhysRevLett.117.190401} {\bibfield  {journal} {\bibinfo  {journal}
  {Phys. Rev. Lett.}\ }\textbf {\bibinfo {volume} {117}},\ \bibinfo {pages}
  {190401} (\bibinfo {year} {2016})}\BibitemShut {NoStop}%
\bibitem [{\citenamefont {Nguyen}\ \emph {et~al.}(2018)\citenamefont {Nguyen},
  \citenamefont {Milne}, \citenamefont {Vu},\ and\ \citenamefont
  {Jevtic}}]{nguyen2018}%
  \BibitemOpen
  \bibfield  {author} {\bibinfo {author} {\bibfnamefont {H.~C.}\ \bibnamefont
  {Nguyen}}, \bibinfo {author} {\bibfnamefont {A.}~\bibnamefont {Milne}},
  \bibinfo {author} {\bibfnamefont {T.}~\bibnamefont {Vu}}, \ and\ \bibinfo
  {author} {\bibfnamefont {S.}~\bibnamefont {Jevtic}},\ }\bibfield  {title}
  {\enquote {\bibinfo {title} {Quantum steering with positive operator valued
  measures},}\ }\href {\doibase 10.1088/1751-8121/aad115} {\bibfield  {journal}
  {\bibinfo  {journal} {J. Phys. A: Math. Theor.}\ }\textbf {\bibinfo {volume}
  {51}},\ \bibinfo {pages} {355302} (\bibinfo {year} {2018})}\BibitemShut
  {NoStop}%
\bibitem [{\citenamefont {Fillettaz}\ \emph {et~al.}(2018)\citenamefont
  {Fillettaz}, \citenamefont {Hirsch}, \citenamefont {Designolle},\ and\
  \citenamefont {Brunner}}]{fillettaz2018}%
  \BibitemOpen
  \bibfield  {author} {\bibinfo {author} {\bibfnamefont {M.}~\bibnamefont
  {Fillettaz}}, \bibinfo {author} {\bibfnamefont {F.}~\bibnamefont {Hirsch}},
  \bibinfo {author} {\bibfnamefont {S.}~\bibnamefont {Designolle}}, \ and\
  \bibinfo {author} {\bibfnamefont {N.}~\bibnamefont {Brunner}},\ }\bibfield
  {title} {\enquote {\bibinfo {title} {Algorithmic construction of local models
  for entangled quantum states: Optimization for two-qubit states},}\ }\href
  {\doibase 10.1103/PhysRevA.98.022115} {\bibfield  {journal} {\bibinfo
  {journal} {Phys. Rev. A}\ }\textbf {\bibinfo {volume} {98}},\ \bibinfo
  {pages} {022115} (\bibinfo {year} {2018})}\BibitemShut {NoStop}%
\bibitem [{\citenamefont {Nguyen}\ \emph {et~al.}(2019)\citenamefont {Nguyen},
  \citenamefont {Nguyen},\ and\ \citenamefont {Gühne}}]{nguyen2019}%
  \BibitemOpen
  \bibfield  {author} {\bibinfo {author} {\bibfnamefont {H.~C.}\ \bibnamefont
  {Nguyen}}, \bibinfo {author} {\bibfnamefont {H.-V.}\ \bibnamefont {Nguyen}},
  \ and\ \bibinfo {author} {\bibfnamefont {O.}~\bibnamefont {Gühne}},\
  }\bibfield  {title} {\enquote {\bibinfo {title} {Geometry of
  {Einstein}-{Podolsky}-{Rosen} correlations},}\ }\href {\doibase
  10.1103/PhysRevLett.122.240401} {\bibfield  {journal} {\bibinfo  {journal}
  {Phys. Rev. Lett.}\ }\textbf {\bibinfo {volume} {122}},\ \bibinfo {pages}
  {240401} (\bibinfo {year} {2019})}\BibitemShut {NoStop}%
\bibitem [{\citenamefont {Tendick}\ \emph {et~al.}(2020)\citenamefont
  {Tendick}, \citenamefont {Kampermann},\ and\ \citenamefont
  {Bruß}}]{activation_bound_entanglement_2020}%
  \BibitemOpen
  \bibfield  {author} {\bibinfo {author} {\bibfnamefont {L.}~\bibnamefont
  {Tendick}}, \bibinfo {author} {\bibfnamefont {H.}~\bibnamefont {Kampermann}},
  \ and\ \bibinfo {author} {\bibfnamefont {D.}~\bibnamefont {Bruß}},\
  }\bibfield  {title} {\enquote {\bibinfo {title} {Activation of nonlocality in
  bound entanglement},}\ }\href {\doibase 10.1103/PhysRevLett.124.050401}
  {\bibfield  {journal} {\bibinfo  {journal} {Phys. Rev. Lett.}\ }\textbf
  {\bibinfo {volume} {124}},\ \bibinfo {pages} {050401} (\bibinfo {year}
  {2020})}\BibitemShut {NoStop}%
\bibitem [{\citenamefont {Villegas-Aguilar}\ \emph {et~al.}(2024)\citenamefont
  {Villegas-Aguilar}, \citenamefont {Polino}, \citenamefont {Ghafari},
  \citenamefont {Quintino}, \citenamefont {Laverick}, \citenamefont {Berkman},
  \citenamefont {Rogge}, \citenamefont {Shalm}, \citenamefont {Tischler},
  \citenamefont {Cavalcanti}, \citenamefont {Slussarenko},\ and\ \citenamefont
  {Pryde}}]{villegasaguilar2023nonlocality}%
  \BibitemOpen
  \bibfield  {author} {\bibinfo {author} {\bibfnamefont {Luis}\ \bibnamefont
  {Villegas-Aguilar}}, \bibinfo {author} {\bibfnamefont {Emanuele}\
  \bibnamefont {Polino}}, \bibinfo {author} {\bibfnamefont {Farzad}\
  \bibnamefont {Ghafari}}, \bibinfo {author} {\bibfnamefont {Marco~Túlio}\
  \bibnamefont {Quintino}}, \bibinfo {author} {\bibfnamefont {Kiarn~T.}\
  \bibnamefont {Laverick}}, \bibinfo {author} {\bibfnamefont {Ian~R.}\
  \bibnamefont {Berkman}}, \bibinfo {author} {\bibfnamefont {Sven}\
  \bibnamefont {Rogge}}, \bibinfo {author} {\bibfnamefont {Lynden~K.}\
  \bibnamefont {Shalm}}, \bibinfo {author} {\bibfnamefont {Nora}\ \bibnamefont
  {Tischler}}, \bibinfo {author} {\bibfnamefont {Eric~G.}\ \bibnamefont
  {Cavalcanti}}, \bibinfo {author} {\bibfnamefont {Sergei}\ \bibnamefont
  {Slussarenko}}, \ and\ \bibinfo {author} {\bibfnamefont {Geoff~J.}\
  \bibnamefont {Pryde}},\ }\bibfield  {title} {\enquote {\bibinfo {title}
  {Nonlocality activation in a photonic quantum network},}\ }\href {\doibase
  10.1038/s41467-024-47354-w} {\bibfield  {journal} {\bibinfo  {journal} {Nat.
  Commun.}\ }\textbf {\bibinfo {volume} {15}},\ \bibinfo {pages} {3112}
  (\bibinfo {year} {2024})}\BibitemShut {NoStop}%
\bibitem [{\citenamefont {Quintino}\ \emph {et~al.}(2016)\citenamefont
  {Quintino}, \citenamefont {Brunner},\ and\ \citenamefont
  {Huber}}]{quintino2016}%
  \BibitemOpen
  \bibfield  {author} {\bibinfo {author} {\bibfnamefont {M.~T.}\ \bibnamefont
  {Quintino}}, \bibinfo {author} {\bibfnamefont {N.}~\bibnamefont {Brunner}}, \
  and\ \bibinfo {author} {\bibfnamefont {M.}~\bibnamefont {Huber}},\ }\bibfield
   {title} {\enquote {\bibinfo {title} {Superactivation of quantum steering},}\
  }\href {\doibase 10.1103/PhysRevA.94.062123} {\bibfield  {journal} {\bibinfo
  {journal} {Phys. Rev. A}\ }\textbf {\bibinfo {volume} {94}},\ \bibinfo
  {pages} {062123} (\bibinfo {year} {2016})}\BibitemShut {NoStop}%
\bibitem [{\citenamefont {Wiseman}\ \emph {et~al.}(2007)\citenamefont
  {Wiseman}, \citenamefont {Jones},\ and\ \citenamefont
  {Doherty}}]{wiseman2007}%
  \BibitemOpen
  \bibfield  {author} {\bibinfo {author} {\bibfnamefont {H.~M.}\ \bibnamefont
  {Wiseman}}, \bibinfo {author} {\bibfnamefont {S.~J.}\ \bibnamefont {Jones}},
  \ and\ \bibinfo {author} {\bibfnamefont {A.~C.}\ \bibnamefont {Doherty}},\
  }\bibfield  {title} {\enquote {\bibinfo {title} {Steering, entanglement,
  nonlocality, and the {Einstein}-{Podolsky}-{Rosen} paradox},}\ }\href
  {\doibase 10.1103/PhysRevLett.98.140402} {\bibfield  {journal} {\bibinfo
  {journal} {Phys. Rev. Lett.}\ }\textbf {\bibinfo {volume} {98}},\ \bibinfo
  {pages} {140402} (\bibinfo {year} {2007})}\BibitemShut {NoStop}%
\bibitem [{\citenamefont {Uola}\ \emph {et~al.}(2020)\citenamefont {Uola},
  \citenamefont {Costa}, \citenamefont {Nguyen},\ and\ \citenamefont
  {Gühne}}]{uola2020}%
  \BibitemOpen
  \bibfield  {author} {\bibinfo {author} {\bibfnamefont {R.}~\bibnamefont
  {Uola}}, \bibinfo {author} {\bibfnamefont {A.~C.~S.}\ \bibnamefont {Costa}},
  \bibinfo {author} {\bibfnamefont {H.~C.}\ \bibnamefont {Nguyen}}, \ and\
  \bibinfo {author} {\bibfnamefont {O.}~\bibnamefont {Gühne}},\ }\bibfield
  {title} {\enquote {\bibinfo {title} {Quantum steering},}\ }\href {\doibase
  10.1103/RevModPhys.92.015001} {\bibfield  {journal} {\bibinfo  {journal}
  {Rev. Mod. Phys.}\ }\textbf {\bibinfo {volume} {92}},\ \bibinfo {pages}
  {015001} (\bibinfo {year} {2020})}\BibitemShut {NoStop}%
\bibitem [{\citenamefont {Tischler}\ \emph {et~al.}(2018)\citenamefont
  {Tischler}, \citenamefont {Ghafari}, \citenamefont {Baker}, \citenamefont
  {Slussarenko}, \citenamefont {Patel}, \citenamefont {Weston}, \citenamefont
  {Wollmann}, \citenamefont {Shalm}, \citenamefont {Verma}, \citenamefont
  {Nam}, \citenamefont {Nguyen}, \citenamefont {Wiseman},\ and\ \citenamefont
  {Pryde}}]{tischler2018}%
  \BibitemOpen
  \bibfield  {author} {\bibinfo {author} {\bibfnamefont {N.}~\bibnamefont
  {Tischler}}, \bibinfo {author} {\bibfnamefont {F.}~\bibnamefont {Ghafari}},
  \bibinfo {author} {\bibfnamefont {T.~J.}\ \bibnamefont {Baker}}, \bibinfo
  {author} {\bibfnamefont {S.}~\bibnamefont {Slussarenko}}, \bibinfo {author}
  {\bibfnamefont {R.~B.}\ \bibnamefont {Patel}}, \bibinfo {author}
  {\bibfnamefont {M.~M.}\ \bibnamefont {Weston}}, \bibinfo {author}
  {\bibfnamefont {S.}~\bibnamefont {Wollmann}}, \bibinfo {author}
  {\bibfnamefont {L.~K.}\ \bibnamefont {Shalm}}, \bibinfo {author}
  {\bibfnamefont {B.~B.}\ \bibnamefont {Verma}}, \bibinfo {author}
  {\bibfnamefont {S.~Woo}\ \bibnamefont {Nam}}, \bibinfo {author}
  {\bibfnamefont {H.~C.}\ \bibnamefont {Nguyen}}, \bibinfo {author}
  {\bibfnamefont {H.~M.}\ \bibnamefont {Wiseman}}, \ and\ \bibinfo {author}
  {\bibfnamefont {G.~J.}\ \bibnamefont {Pryde}},\ }\bibfield  {title} {\enquote
  {\bibinfo {title} {Conclusive experimental demonstration of one-way
  {Einstein}-{Podolsky}-{Rosen} steering},}\ }\href {\doibase
  10.1103/PhysRevLett.121.100401} {\bibfield  {journal} {\bibinfo  {journal}
  {Phys. Rev. Lett.}\ }\textbf {\bibinfo {volume} {121}},\ \bibinfo {pages}
  {100401} (\bibinfo {year} {2018})}\BibitemShut {NoStop}%
\bibitem [{\citenamefont {Werner}(1989)}]{werner1989}%
  \BibitemOpen
  \bibfield  {author} {\bibinfo {author} {\bibfnamefont {R.~F.}\ \bibnamefont
  {Werner}},\ }\bibfield  {title} {\enquote {\bibinfo {title} {Quantum states
  with {Einstein}-{Podolsky}-{Rosen} correlations admitting a hidden-variable
  model},}\ }\href {\doibase 10.1103/PhysRevA.40.4277} {\bibfield  {journal}
  {\bibinfo  {journal} {Phys. Rev. A}\ }\textbf {\bibinfo {volume} {40}},\
  \bibinfo {pages} {4277} (\bibinfo {year} {1989})}\BibitemShut {NoStop}%
\bibitem [{\citenamefont {V\'ertesi}\ and\ \citenamefont
  {Bene}(2010)}]{Vertesi2010_TwoQubitBell}%
  \BibitemOpen
  \bibfield  {author} {\bibinfo {author} {\bibfnamefont {T.}~\bibnamefont
  {V\'ertesi}}\ and\ \bibinfo {author} {\bibfnamefont {E.}~\bibnamefont
  {Bene}},\ }\bibfield  {title} {\enquote {\bibinfo {title} {Two-qubit {Bell}
  inequality for which positive operator-valued measurements are relevant},}\
  }\href {\doibase 10.1103/PhysRevA.82.062115} {\bibfield  {journal} {\bibinfo
  {journal} {Phys. Rev. A}\ }\textbf {\bibinfo {volume} {82}},\ \bibinfo
  {pages} {062115} (\bibinfo {year} {2010})}\BibitemShut {NoStop}%
\bibitem [{\citenamefont {Evans}\ \emph {et~al.}(2013)\citenamefont {Evans},
  \citenamefont {Cavalcanti},\ and\ \citenamefont {Wiseman}}]{evans2013}%
  \BibitemOpen
  \bibfield  {author} {\bibinfo {author} {\bibfnamefont {D.~A.}\ \bibnamefont
  {Evans}}, \bibinfo {author} {\bibfnamefont {E.~G.}\ \bibnamefont
  {Cavalcanti}}, \ and\ \bibinfo {author} {\bibfnamefont {H.~M.}\ \bibnamefont
  {Wiseman}},\ }\bibfield  {title} {\enquote {\bibinfo {title} {Loss-tolerant
  tests of {Einstein}-{Podolsky}-{Rosen} steering},}\ }\href {\doibase
  10.1103/PhysRevA.88.022106} {\bibfield  {journal} {\bibinfo  {journal} {Phys.
  Rev. A}\ }\textbf {\bibinfo {volume} {88}},\ \bibinfo {pages} {022106}
  (\bibinfo {year} {2013})}\BibitemShut {NoStop}%
\bibitem [{\citenamefont {Baker}\ \emph {et~al.}(2018)\citenamefont {Baker},
  \citenamefont {Wollmann}, \citenamefont {Pryde},\ and\ \citenamefont
  {Wiseman}}]{baker2018}%
  \BibitemOpen
  \bibfield  {author} {\bibinfo {author} {\bibfnamefont {T.~J.}\ \bibnamefont
  {Baker}}, \bibinfo {author} {\bibfnamefont {S.}~\bibnamefont {Wollmann}},
  \bibinfo {author} {\bibfnamefont {G.~J.}\ \bibnamefont {Pryde}}, \ and\
  \bibinfo {author} {\bibfnamefont {H.~M.}\ \bibnamefont {Wiseman}},\
  }\bibfield  {title} {\enquote {\bibinfo {title} {Necessary condition for
  steerability of arbitrary two-qubit states with loss},}\ }\href {\doibase
  10.1088/2040-8986/aaaa3c} {\bibfield  {journal} {\bibinfo  {journal} {J.
  Opt.}\ }\textbf {\bibinfo {volume} {20}},\ \bibinfo {pages} {034008}
  (\bibinfo {year} {2018})}\BibitemShut {NoStop}%
\bibitem [{\citenamefont {Quintino}\ \emph {et~al.}(2015)\citenamefont
  {Quintino}, \citenamefont {Vértesi}, \citenamefont {Cavalcanti},
  \citenamefont {Augusiak}, \citenamefont {Demianowicz}, \citenamefont
  {Acín},\ and\ \citenamefont {Brunner}}]{quintino2015}%
  \BibitemOpen
  \bibfield  {author} {\bibinfo {author} {\bibfnamefont {M.~T.}\ \bibnamefont
  {Quintino}}, \bibinfo {author} {\bibfnamefont {T.}~\bibnamefont {Vértesi}},
  \bibinfo {author} {\bibfnamefont {D.}~\bibnamefont {Cavalcanti}}, \bibinfo
  {author} {\bibfnamefont {R.}~\bibnamefont {Augusiak}}, \bibinfo {author}
  {\bibfnamefont {M.}~\bibnamefont {Demianowicz}}, \bibinfo {author}
  {\bibfnamefont {A.}~\bibnamefont {Acín}}, \ and\ \bibinfo {author}
  {\bibfnamefont {N.}~\bibnamefont {Brunner}},\ }\bibfield  {title} {\enquote
  {\bibinfo {title} {Inequivalence of entanglement, steering, and {Bell}
  nonlocality for general measurements},}\ }\href {\doibase
  10.1103/PhysRevA.92.032107} {\bibfield  {journal} {\bibinfo  {journal} {Phys.
  Rev. A}\ }\textbf {\bibinfo {volume} {92}},\ \bibinfo {pages} {032107}
  (\bibinfo {year} {2015})}\BibitemShut {NoStop}%
\bibitem [{\citenamefont {Barrett}(2002)}]{barrett2002}%
  \BibitemOpen
  \bibfield  {author} {\bibinfo {author} {\bibfnamefont {J.}~\bibnamefont
  {Barrett}},\ }\bibfield  {title} {\enquote {\bibinfo {title} {Nonsequential
  positive-operator-valued measurements on entangled mixed states do not always
  violate a {B}ell inequality},}\ }\href {\doibase 10.1103/PhysRevA.65.042302}
  {\bibfield  {journal} {\bibinfo  {journal} {Phys. Rev. A}\ }\textbf {\bibinfo
  {volume} {65}},\ \bibinfo {pages} {042302} (\bibinfo {year}
  {2002})}\BibitemShut {NoStop}%
\bibitem [{\citenamefont {Clauser}\ \emph {et~al.}(1969)\citenamefont
  {Clauser}, \citenamefont {Horne}, \citenamefont {Shimony},\ and\
  \citenamefont {Holt}}]{chsh_1969}%
  \BibitemOpen
  \bibfield  {author} {\bibinfo {author} {\bibfnamefont {J.~F.}\ \bibnamefont
  {Clauser}}, \bibinfo {author} {\bibfnamefont {M.~A.}\ \bibnamefont {Horne}},
  \bibinfo {author} {\bibfnamefont {A.}~\bibnamefont {Shimony}}, \ and\
  \bibinfo {author} {\bibfnamefont {R.~A.}\ \bibnamefont {Holt}},\ }\bibfield
  {title} {\enquote {\bibinfo {title} {Proposed experiment to test local
  hidden-variable theories},}\ }\href {\doibase 10.1103/PhysRevLett.23.880}
  {\bibfield  {journal} {\bibinfo  {journal} {Phys. Rev. Lett.}\ }\textbf
  {\bibinfo {volume} {23}},\ \bibinfo {pages} {880} (\bibinfo {year}
  {1969})}\BibitemShut {NoStop}%
\bibitem [{\citenamefont {Amselem}\ and\ \citenamefont
  {Bourennane}(2009)}]{amselem_experimental_2009}%
  \BibitemOpen
  \bibfield  {author} {\bibinfo {author} {\bibfnamefont {Elias}\ \bibnamefont
  {Amselem}}\ and\ \bibinfo {author} {\bibfnamefont {Mohamed}\ \bibnamefont
  {Bourennane}},\ }\bibfield  {title} {\enquote {\bibinfo {title} {Experimental
  four-qubit bound entanglement},}\ }\href {\doibase 10.1038/nphys1372}
  {\bibfield  {journal} {\bibinfo  {journal} {Nature Phys.}\ }\textbf {\bibinfo
  {volume} {5}},\ \bibinfo {pages} {748} (\bibinfo {year} {2009})}\BibitemShut
  {NoStop}%
\bibitem [{\citenamefont {Lavoie}\ \emph
  {et~al.}(2010{\natexlab{a}})\citenamefont {Lavoie}, \citenamefont
  {Kaltenbaek}, \citenamefont {Piani},\ and\ \citenamefont
  {Resch}}]{lavoie_experimental_2010}%
  \BibitemOpen
  \bibfield  {author} {\bibinfo {author} {\bibfnamefont {J.}~\bibnamefont
  {Lavoie}}, \bibinfo {author} {\bibfnamefont {R.}~\bibnamefont {Kaltenbaek}},
  \bibinfo {author} {\bibfnamefont {M.}~\bibnamefont {Piani}}, \ and\ \bibinfo
  {author} {\bibfnamefont {K.~J.}\ \bibnamefont {Resch}},\ }\bibfield  {title}
  {\enquote {\bibinfo {title} {Experimental bound entanglement?}}\ }\href
  {\doibase 10.1038/nphys1832} {\bibfield  {journal} {\bibinfo  {journal}
  {Nature Phys.}\ }\textbf {\bibinfo {volume} {6}},\ \bibinfo {pages} {827}
  (\bibinfo {year} {2010}{\natexlab{a}})}\BibitemShut {NoStop}%
\bibitem [{\citenamefont {Lavoie}\ \emph
  {et~al.}(2010{\natexlab{b}})\citenamefont {Lavoie}, \citenamefont
  {Kaltenbaek}, \citenamefont {Piani},\ and\ \citenamefont
  {Resch}}]{lavoie_experimental_2010-1}%
  \BibitemOpen
  \bibfield  {author} {\bibinfo {author} {\bibfnamefont {Jonathan}\
  \bibnamefont {Lavoie}}, \bibinfo {author} {\bibfnamefont {Rainer}\
  \bibnamefont {Kaltenbaek}}, \bibinfo {author} {\bibfnamefont {Marco}\
  \bibnamefont {Piani}}, \ and\ \bibinfo {author} {\bibfnamefont {Kevin~J.}\
  \bibnamefont {Resch}},\ }\bibfield  {title} {\enquote {\bibinfo {title}
  {Experimental {Bound} {Entanglement} in a {Four}-{Photon} {State}},}\ }\href
  {\doibase 10.1103/PhysRevLett.105.130501} {\bibfield  {journal} {\bibinfo
  {journal} {Phys. Rev. Lett.}\ }\textbf {\bibinfo {volume} {105}},\ \bibinfo
  {pages} {130501} (\bibinfo {year} {2010}{\natexlab{b}})}\BibitemShut
  {NoStop}%
\bibitem [{\citenamefont {Moroder}\ \emph {et~al.}(2014)\citenamefont
  {Moroder}, \citenamefont {Gittsovich}, \citenamefont {Huber},\ and\
  \citenamefont {G\"uhne}}]{Moroder2014}%
  \BibitemOpen
  \bibfield  {author} {\bibinfo {author} {\bibfnamefont {Tobias}\ \bibnamefont
  {Moroder}}, \bibinfo {author} {\bibfnamefont {Oleg}\ \bibnamefont
  {Gittsovich}}, \bibinfo {author} {\bibfnamefont {Marcus}\ \bibnamefont
  {Huber}}, \ and\ \bibinfo {author} {\bibfnamefont {Otfried}\ \bibnamefont
  {G\"uhne}},\ }\bibfield  {title} {\enquote {\bibinfo {title} {Steering bound
  entangled states: A counterexample to the stronger {Peres} conjecture},}\
  }\href {\doibase 10.1103/PhysRevLett.113.050404} {\bibfield  {journal}
  {\bibinfo  {journal} {Phys. Rev. Lett.}\ }\textbf {\bibinfo {volume} {113}},\
  \bibinfo {pages} {050404} (\bibinfo {year} {2014})}\BibitemShut {NoStop}%
\bibitem [{\citenamefont {V\'ertesi}\ and\ \citenamefont
  {Brunner}(2014)}]{Vertesi2014}%
  \BibitemOpen
  \bibfield  {author} {\bibinfo {author} {\bibfnamefont {Tam\'as}\ \bibnamefont
  {V\'ertesi}}\ and\ \bibinfo {author} {\bibfnamefont {Nicolas}\ \bibnamefont
  {Brunner}},\ }\bibfield  {title} {\enquote {\bibinfo {title} {Disproving the
  {Peres} conjecture by showing {Bell} nonlocality from bound entanglement},}\
  }\href {\doibase 10.1038/ncomms6297} {\bibfield  {journal} {\bibinfo
  {journal} {Nat. Commun.}\ }\textbf {\bibinfo {volume} {5}},\ \bibinfo {pages}
  {5297} (\bibinfo {year} {2014})}\BibitemShut {NoStop}%
\bibitem [{\citenamefont {Hirsch}\ \emph {et~al.}(2017)\citenamefont {Hirsch},
  \citenamefont {Quintino}, \citenamefont {Vértesi}, \citenamefont
  {Navascués},\ and\ \citenamefont {Brunner}}]{LHVmodels_Werner_2017}%
  \BibitemOpen
  \bibfield  {author} {\bibinfo {author} {\bibfnamefont {F.}~\bibnamefont
  {Hirsch}}, \bibinfo {author} {\bibfnamefont {M.~T.}\ \bibnamefont
  {Quintino}}, \bibinfo {author} {\bibfnamefont {T.}~\bibnamefont {Vértesi}},
  \bibinfo {author} {\bibfnamefont {M.}~\bibnamefont {Navascués}}, \ and\
  \bibinfo {author} {\bibfnamefont {N.}~\bibnamefont {Brunner}},\ }\bibfield
  {title} {\enquote {\bibinfo {title} {Better local hidden variable models for
  two-qubit {Werner} states and an upper bound on the {Grothendieck} constant
  {$K_{G}(3)$}},}\ }\href {\doibase 10.22331/q-2017-04-25-3} {\bibfield
  {journal} {\bibinfo  {journal} {Quantum}\ }\textbf {\bibinfo {volume} {1}},\
  \bibinfo {pages} {3} (\bibinfo {year} {2017})}\BibitemShut {NoStop}%
\end{thebibliography}%

\end{document}